\documentclass[aps,twocolumn,showpacs,superscriptaddress,floats,prb]{revtex4}

\usepackage{amsmath}
\usepackage{graphicx,epsfig,psfrag}
\usepackage{amssymb}

\newcommand{\nn}{\nonumber}

\newcommand{\Tcl}{T_{cl}}

\newcommand{\K}{\mathbf{k}}

\newcommand{\br}{\mathbf{r}}

\newcommand{\me}{e}

\newcommand{\TN}{T_{\rm N}}
\newcommand{\Tc}{T_{\rm c}}

\newcommand{\CeAu}{CeCu$_{6-x}$Au$_x$}

\newcommand{\YbRhSi}{YbRh$_2$Si$_2$}
\newcommand{\YbRhSG}{YbRh$_2$(Si$_{1-x}$Ge$_x$)$_2$}
\newcommand{\CeCoIn}{CeCoIn$_5$}

\begin{document}

\title{
Dimensional crossover in quantum critical metallic magnets
}

\author{Markus Garst}
\affiliation{Institut f\"ur Theoretische Physik, Universit\"at zu K\"oln,
Z\"ulpicher Str. 77, 50937 K\"oln, Germany
}
\author{Lars Fritz}
\affiliation{Department of Physics, Harvard University, Cambridge MA 02138, USA}
\author{Achim Rosch}
\author{Matthias Vojta}
\affiliation{Institut f\"ur Theoretische Physik, Universit\"at zu K\"oln,
Z\"ulpicher Str. 77, 50937 K\"oln, Germany
}

\begin{abstract}
Nearly magnetic metals often have layered lattice structures, consisting of
coupled planes. In such a situation, physical properties will display,
upon decreasing temperature or energy, a dimensional crossover from
two-dimensional (2d) to three-dimensional (3d) behavior, which is
particularly interesting near quantum criticality.
Here we study this crossover in thermodynamics
using a suitably generalized Landau-Ginzburg-Wilson approach to the critical behavior,
combined with renormalization group techniques.
We focus on two experimentally relevant cases: the crossover from a 2d to a 3d
antiferromagnet, and the crossover from a 2d ferromagnet to a 3d antiferromagnet.
As naive scaling does not apply at and above the upper critical dimension,
two crossover scales arise which can be associated with separate dimensional crossovers
of classical and quantum fluctuations, respectively. In particular, we find an intermediate regime with novel power laws where the quantum fluctuations still have a 2d and the classical fluctuations already have a 3d character.
For the ferromagnet-to-antiferromagnet crossover, the mismatch of the dynamical
exponents between the 2d and 3d regimes leads to an even richer crossover structure,
with an interesting 2d non-critical regime sandwiched between two critical regimes.
For all cases, we find that thermal expansion and compressibility are particularly
sensitive probes of the dimensional crossover.
Finally, we relate our results to experiments on the quantum critical
heavy-fermion metals \CeAu, \YbRhSi, and \CeCoIn.
\end{abstract}

\date{\today}

\maketitle


\section{Introduction}

Quantum phase transitions (QPT) in metals are a fascinating field of today's condensed
matter research.\cite{hvl} Heavy-fermion materials play a prominent role: Frequently, one
observes non-Fermi liquid behavior which is thought to be associated with an
antiferromagnetic (AFM) instability of the itinerant electrons. The critical spin
fluctuations near the phase transition lead to unconventional power laws in transport and
thermodynamic quantities at low temperatures. On the theory side, a
Landau-Ginzburg-Wilson (LGW) description of the critical magnetic degrees of freedom,
developed by Hertz,\cite{hertz} Moriya,\cite{moriya}, and Millis\cite{millis} accounts
for many of the experimental signatures of magnetic criticality.

However, some heavy-fermion compounds do not easily fit in the LGW picture in
$d=3$ spatial dimensions.
For \CeAu, it was realized\cite{Rosch97} that most thermodynamic signatures
of the QPT at $x=0.1$ are consistent with the assumption of the underlying
AFM spin fluctuations to be 2d. It came as a surprise that
2d spin fluctuations should prevail in an intrinsically
3d alloy, but the 2d character was subsequently
confirmed in neutron-scattering experiments.\cite{Stockert98}
Importantly, the AFM order observed in \CeAu\ below the N\'eel
temperature is fully 3d.
On general grounds, one thus expects a dimensional crossover within the paramagnetic phase
from 2d magnetic fluctuations at elevated temperatures to 3d fluctuations
at lowest temperatures or in the immediate vicinity of the phase
transition.
Experimentally, the dimensional crossover in \CeAu\ has proven to be
elusive so far.

A related heavy-fermion metal, not easily fitting the LGW theory framework,
is \YbRhSi.\cite{trovarelli}
It shows a phase transition to an ordered phase at 70~mK, which is
believed to be AFM, however, a confirmation by neutron scattering is not available
to date.
An additional aspect is that \YbRhSi\ seems to be almost ferromagnetic (FM),\cite{gegenwart05}
and we will return to this issue later in this paper.
The unusual properties of both \CeAu\ and \YbRhSi\ have prompted speculations
on the inapplicability of the LGW theory, which describes a magnetic instability
of well-defined quasiparticles:
Instead, it was proposed that the Kondo effect,
being responsible for the formation of the heavy quasiparticles,
breaks down at the quantum critical point (QCP).\cite{si,coleman}
Different scenarios and theoretical descriptions of this
Kondo breakdown have been put forward.\cite{si,coleman,senthil}
The scenario of so-called ``local quantum criticality''\cite{si}
uses an extension of dynamical mean-field theory to map the Kondo-lattice problem
to a self-consistent impurity model where the Kondo effect may be suppressed
by critical bulk spin fluctuations.
This particular scenario for a Kondo-breakdown QCP requires
the spin fluctuations to be 2d; for 3d spin fluctuations this model predicts
a conventional magnetic QCP of LGW type.
As the critical spin fluctuations of the material can again be expected
to become 3d at low energies, the local quantum criticality should be restricted
to elevated temperatures or energies above the dimensional crossover.

Other examples of layered metals with magnetic QCP are the
heavy fermions CeMIn$_5$ (M=Co,Rh,Ir), the high-temperature superconducting
cuprates and iron pnictides, and the metamagnetic ruthenate Sr$_2$Ru$_3$O$_7$.

The purpose of this paper is to study theoretically the dimensional crossover of
critical magnetic fluctuations in the framework of the LGW model, with
focus on the 2d AFM to 3d AFM and 2d FM to 3d AFM crossovers.
Microscopically, we imagine a system of planes of interacting electrons,
with tendency toward antiferromagnetic or ferromagnetic in-plane ordering,
and a weak antiferromagnetic inter-plane coupling.
We will study how the dimensional crossover is reflected in the
correlation length, specific heat, thermal expansion, compressibility,
and the Gr\"uneisen parameter.\cite{Zhu03}
Our primary goal is to identify observables which are suited for
an experimental search for a dimensional crossover.

In this paper, we shall refrain from a detailed microscopic modeling
of the magnetic inter-plane coupling. The underlying lattice geometry and
bandstructure will influence some of the non-universal properties of the
dimensional crossover; however, the existence of well-defined 2d and 3d regimes
is independent of those details.
We shall ignore complications arising from the possibility of the inter-plane
coupling being frustrated:\cite{sebastian,colemanxover,ormv}
Even for a fully frustrated coupling, a dimensional crossover to 3d behavior
at low energies will generically occur, albeit with a possibly small crossover
scale.\cite{colemanxover,ormv}
(In situations with frustration, the effective 3d coupling within
the ordered phase may be enhanced due to order-from-disorder mechanisms.)

We shall restrict our analysis to the framework of the LGW theory of itinerant
spin fluctuations.
It has been discussed that this approach may break down at lowest energies
due to the occurrence of singular terms in the LGW expansion
(i) for ferromagnets in both 2d and 3d,\cite{bkv} and
(ii) for 2d antiferromagnets.\cite{chub2dafm}
These complications will be ignored for simplicity, a justification being that
the low-energy behavior of our model is invariably 3d AFM
(where the LGW approach is believed to be valid).\cite{intnote}
Our theory may also be combined with the ideas of
local quantum criticality,\cite{si}
this is beyond the scope of this paper.
Similarly, an explicit treatment of the magnetically ordered phases
shall not be performed here.\cite{metzner,afnote}

\subsection{Summary of results}

\begin{figure}
\includegraphics[width=0.47\textwidth]{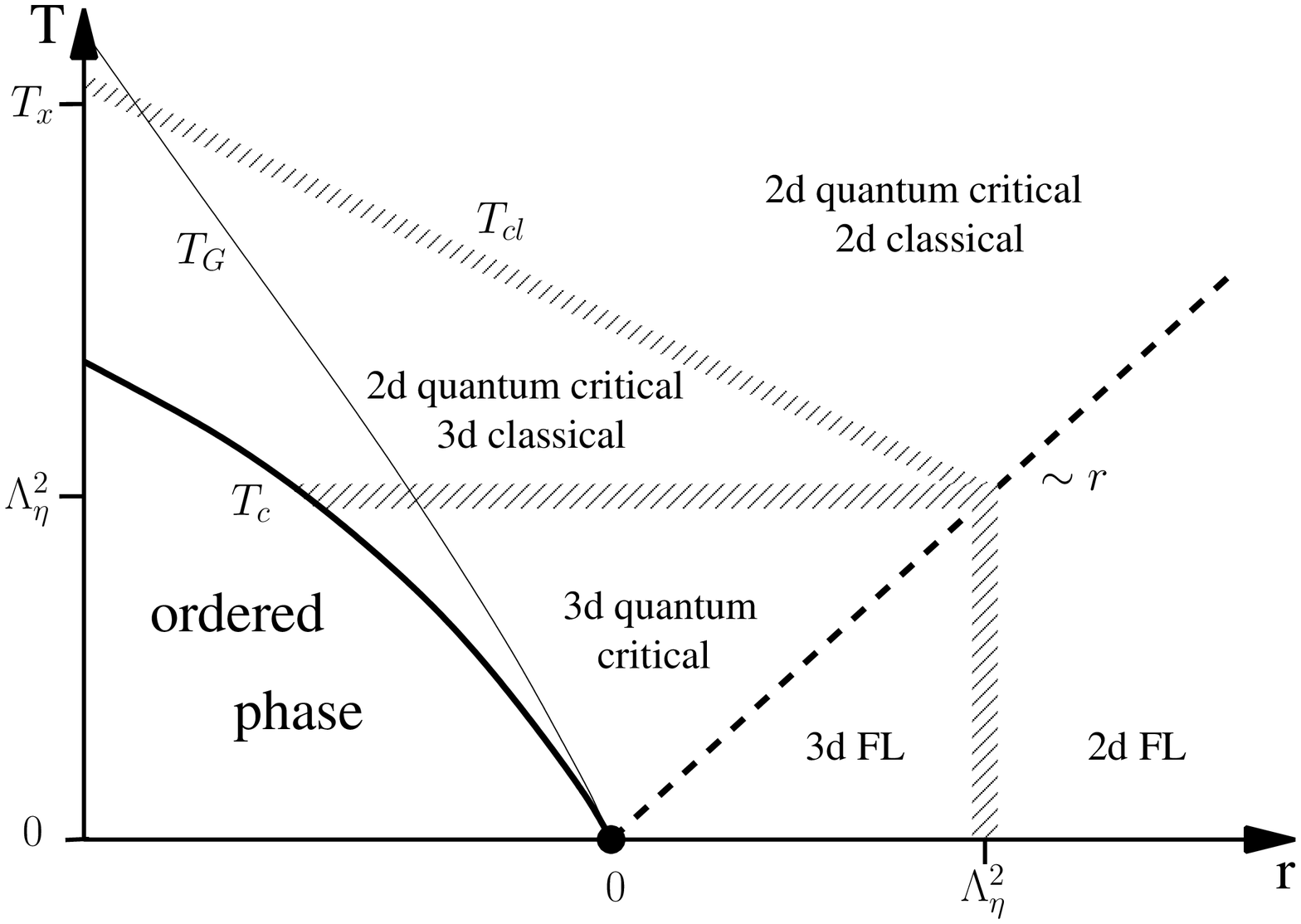}
\caption{
Phase diagram of the anisotropic LGW model in the temperature--control
parameter plane for the 2d AFM to 3d AFM crossover. The crossover scale is determined by
the momentum scale $\Lambda_\eta$. The 3d spin fluctuations dominate close to the quantum
critical point. A dimensional crossover occurs upon increasing the distance to the QCP
indicated by the shaded region.
The phase boundary $T_c(r)$ changes its behavior at this crossover, see
Fig.~\ref{fig:asd} below.
There is an additional dimensional crossover at the temperature scale $\Tcl(r)$ where $\xi
\sim 1/\Lambda_\eta$, associated with the classical critical fluctuations. The thin line,
$T_G(r)$, close to the critical temperature, $T_c(r)$, indicates the Ginzburg temperature
where the crossover to classical Wilson-Fisher behavior occurs. The two lines $\Tcl(r)$ and $T_G(r)$ cross at a temperature $T_x$. The dashed line separates
the low-temperature magnetically disordered (Fermi-liquid, FL) regime, $T \ll r$, from
the quantum critical regime, $T \gg r$.
Note that the labels ``2d'' and ``3d'' refer to the behavior of the
critical or near-critical spin fluctuations, for details see text.
}
\label{fig:pd-AFAF}
\end{figure}

\begin{figure}
\includegraphics[width=0.47\textwidth]{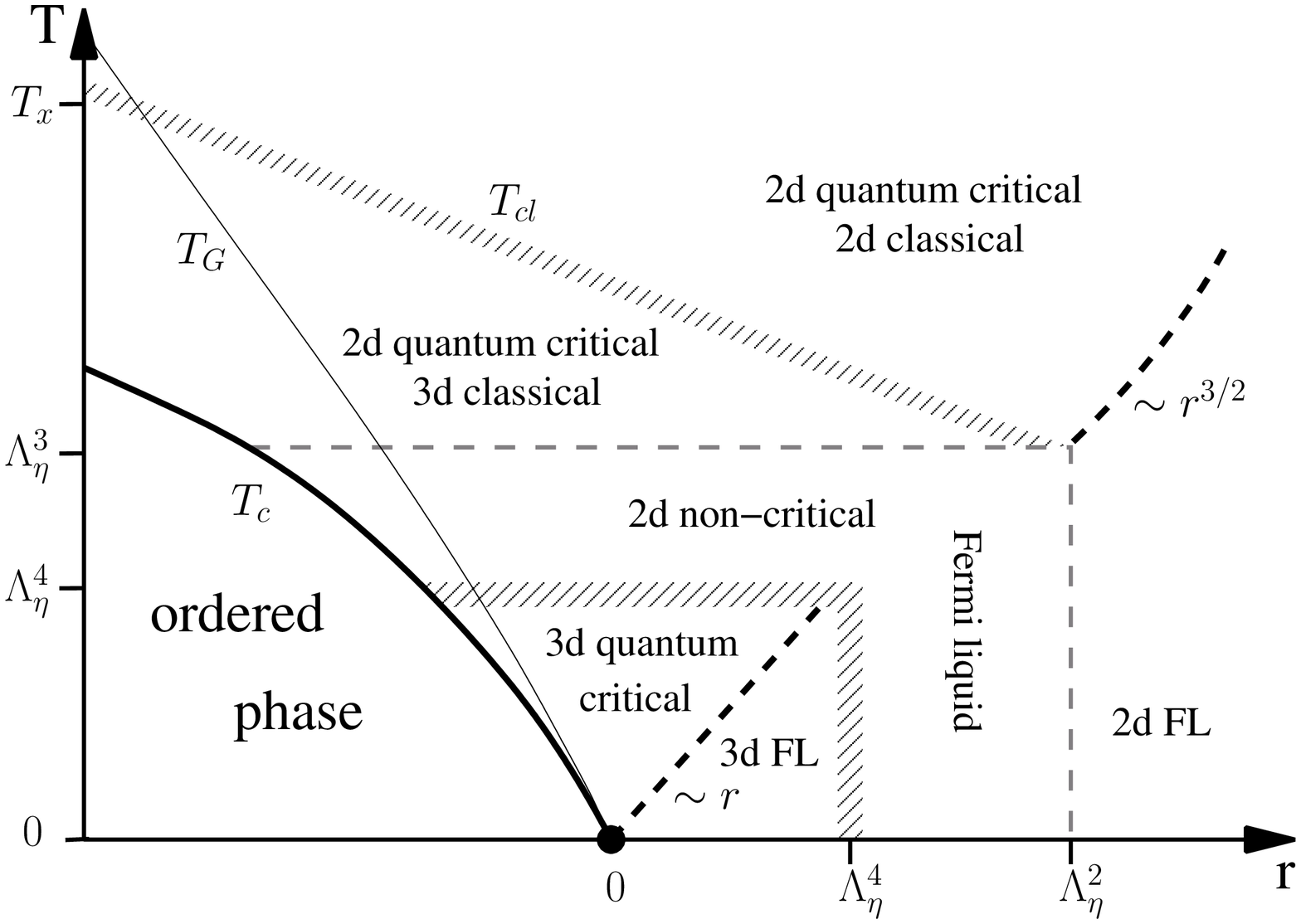}
\caption{
Phase diagram of the anisotropic LGW model as in Fig.~\ref{fig:pd-AFAF}, but for
the 2d FM to 3d AFM crossover. Here, the critical 2d regime, $T > \Lambda^3_\eta$ or $r >
\Lambda^2_\eta$, is separated from the critical 3d regime, $T< \Lambda_\eta^4$ and
$r<\Lambda_\eta^4$, by a region where 2d non-critical Fermi liquid behavior prevails.
}
\label{fig:pd-FMAF}
\end{figure}

The main results of our analysis are the crossover phase diagrams,
Figs.~\ref{fig:pd-AFAF} and \ref{fig:pd-FMAF},
for the 2d -- 3d AFM and the 2d FM -- 3d AFM crossover, respectively.
The large anisotropy in the spin-fluctuation spectrum defines a small momentum scale
$\Lambda_{\eta}$ [see Eq.~\eqref{CrossoverMomentum} below]
that determines the positions of the dimensional crossovers in the phase
diagram, as indicated by the shaded areas.
Generally, the behavior changes from 2d to 3d upon approaching criticality.

However, the fact that the QPT under consideration are at or above their upper critical
dimension renders naive scaling invalid; as a result, the quantum critical regime is
characterized by {\em two} distinct length scales:
The physical correlation length $\xi$ and a thermal length given by $T^{-1/z}$,
where $T$ is the temperature and $z$ the dynamical exponent.
The presence of two length scales results in two types of dimensional crossovers.

\subsubsection{Quantum crossover}

A dimensional crossover in the quantum critical fluctuations occurs upon
approaching the QCP either by lowering the temperature $T$ or decreasing
the tuning parameter $r$;
these crossovers are indicated by the horizontal and vertical
shaded regions in Figs.~\ref{fig:pd-AFAF} and \ref{fig:pd-FMAF}.

In the case of the 2d to 3d AF crossover, the positions of these crossover lines are given
by $T \sim \Lambda_\eta^z$ and $r \sim \Lambda_\eta^{1/\nu}$, where $z=2$ and the
correlation-length exponent has the mean-field value $\nu = 1/2$.
For example, the specific heat coefficient at criticality, $r=0$, changes its temperature
dependence at the crossover temperature, $T \sim \Lambda_\eta^2$, from $\gamma \sim
\log(1/T)$ to $\gamma \sim {\rm const} - \sqrt{T}$, as expected for critical
thermodynamics of 2d and 3d AF fluctuations, respectively.

The situation is more complicated for the 2d FM to 3d AFM crossover, due to the mismatch of
dynamical exponents in the two regimes. The low dimensionality, $d=2$, of the FM
spin fluctuations combined with a large dynamical exponent, $z=3$,
result in strong thermodynamic signatures of the 2d regime,
which dominate over the 3d AFM fluctuations in an unexpectedly wide regime.
For example, at criticality, $r=0$, the 2d FM fluctuations yield singular quantum
critical thermodynamics down to a temperature scale $T \sim \Lambda^z_\eta$, where $z=3$.
Below this temperature scale, the thermal activation of these 2d fluctuations only yields
Fermi-liquid behavior, but this is still much stronger than the contribution
from the 3d part of the spin-fluctuation spectrum.
Only at a much lower temperature scale, $T \sim \Lambda_\eta^4$,
the contributions from the 3d AFM fluctuations finally take over.
Thus, we have here the peculiar situation that a 2d non-critical Fermi-liquid regime
is sandwiched between the 2d and 3d quantum critical regions.

\subsubsection{Classical crossover}

In addition, there is a dimensional crossover associated with classical criticality.
Upon approaching the classical phase transition line, $T_c(r)$,
the correlation length $\xi$ increases.
If $\xi$ reaches $1/\Lambda_\eta$, the classical fluctuations
(i.e. those associated with zero Matsubara frequency) effectively change
their dimensionality from 2d to 3d.
This classical dimensional crossover occurs {\em within} the 2d quantum critical regime
at the shaded line labeled $\Tcl(r)$ in Figs.~\ref{fig:pd-AFAF} and \ref{fig:pd-FMAF},
and it also causes thermodynamic signatures.
For example, the thermal expansion $\alpha(T)$ has a maximum at $\Tcl$
for the 2d FM to 3d AFM crossover. At higher temperature $T_x$, the crossover line $\Tcl$ enters the Ginzburg regime of the classical transition, and the classical dimensional crossover becomes non-perturbative.

\subsubsection{Phase boundary and QCP location}

The phase boundary, $T_c(r)$, of the classical transition is linear in the distance to
the QCP at elevated temperatures (with logarithmic corrections), but curves towards the
QCP in the 3d regime. As a consequence, an extrapolation of the
quasi-linear phase boundary in the 2d regime towards zero temperature yields an
incorrect position for the QCP, see Fig.~\ref{fig:asd},
and we estimate the corresponding error between the extrapolated and the true
position of the QPT.

We note that our treatment of thermodynamics is limited to the non-ordered side
of the phase transition and, in particular, breaks down upon entering the Ginzburg regime
of the classical critical transition indicated by the thin line $T_G(r)$ in
Figs.~\ref{fig:pd-AFAF} and \ref{fig:pd-FMAF}.

\subsubsection{Observables}

As detailed below, we find that the thermal expansion and the compressibility are well suited to
detect a dimensional crossover in the spin-fluctuation spectrum. Both possess pronounced
signatures close to the expected crossovers as a function of temperature in the quantum
critical regime, either a sharp drop or even a maximum. In contrast to this, the specific
heat coefficient only shows a leveling off upon entering the 3d regime,
which is harder to identify experimentally,
see Figs.~\ref{fig:2af3af} and \ref{fig:2f3af} below.

\subsection{Outline}

The body of the paper is organized as follows:
In Sec.~\ref{sec:action} we introduce the Landau-Ginzburg-Wilson field theory for
magnetism near quantum criticality.
We discuss the anisotropic spin susceptibility and the associated crossover
in the Landau damping. We introduce our model for the dimensional crossover and give the
resulting formulae that determine the correlation length and other thermodynamic
properties.
Secs.~\ref{sec:2af3af} and \ref{sec:2f3af} are devoted to a detailed discussion
of the 2d -- 3d AFM and 2d FM -- 3d AFM crossovers, respectively.
We shall derive phase diagrams and full crossover functions for thermodynamic
quantities.
Finally, in Sec.~\ref{sec:exp} we discuss existing experimental
data vis-a-vis our theory results.
We focus on the heavy-fermion metals \CeAu, \YbRhSi, and \CeCoIn\ which indeed display
unconventional quantum criticality that may originate from quasi-2d
spin fluctuations.
A brief outlook concludes the paper.


\section{Order-parameter field theory for spatially anisotropic spin fluctuations}
\label{sec:action}

In order to analyze the dimensional crossover we will use the standard LGW
critical theory of Hertz, Millis and Moriya for a (commensurate) itinerant paramagnet. The
action of the Hertz-Millis-Moriya model reads\cite{hvl,hertz,millis}
\begin{align}\label{HertzModel}
\mathcal{S}  &=
 \int_0^\beta d\tau \int d^d\br \left[
\frac{1}{2} \Phi^T \chi^{-1}_0( - i \nabla,\partial_\tau) \Phi
+ \frac{u_0}{4!}
\left( \Phi^T \Phi \right)^2
\right].
\end{align}
where the real bosonic order-parameter field $\Phi$ represents commensurate spin fluctuations
with a 3d ordering wavevector $\bf Q$.
We will generalize the field $\Phi$ to have $N$ components; the Heisenberg paramagnet
corresponds to $N=3$.
The dynamics of the fluctuations are encoded in the propagator $\chi^{-1}_0$. Its
momentum dependence will reflect the spatial anisotropy of the spin-fluctuation spectrum.
Its form will be motivated in the following.

\subsection{Bare susceptibility}
\label{sec:bare}

\begin{figure}
\includegraphics[width=0.48\textwidth,clip=true]{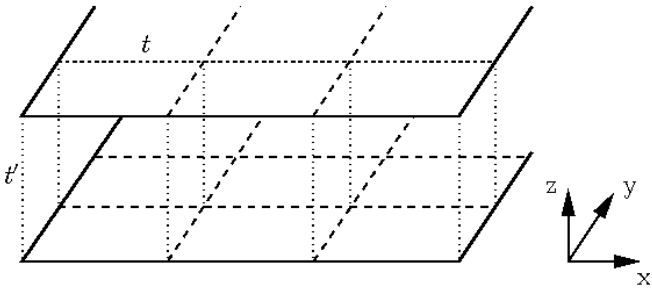}
\caption{
\label{fig:layer}
Schematics of a layered crystal structure with a weak hopping of spin-fluctuations between the planes, $t' \ll t$.
}
\end{figure}

Starting from a model of interacting electrons on a 3d anisotropic lattice,
a Hubbard-Stratonovich transformation allows to introduce collective-mode
variables representing the spin-fluctuations. Their dispersion can be estimated, e.g., by RPA.
Consider for simplicity a paramagnon dispersion of tight-binding type.
On a 3d tetragonal lattice, a generic form is
\begin{equation} \label{dispersion}
\omega(\K) =
t \left(2 - \cos k_x a - \cos k_y a \right) +
t'\left(1 - \cos k_z a \right)
\end{equation}
where $t$ and $t'$ parameterize the hopping of the spin fluctuations
within and perpendicular to the $xy$ planes, see Fig.~\ref{fig:layer},
the momentum $\K$ is measured relative to the ordering wavevector $\bf Q$,
and $a$ is a lattice constant.
The quasi-2d character of the spin fluctuations is reflected in a small ratio
between the hopping amplitudes,
\begin{equation}
\eta \equiv t'/t \ll 1.
\end{equation}

Near the transition we can restrict ourselves to energies much smaller than the (large)
in-plane bandwidth $\propto t$; this is equivalent to a continuum approximation
w.r.t. the in-plane lattice coordinates. The result is
\begin{equation}
\omega(\K)= t \xi_0^2 \left (k_\parallel^2+\frac{2\eta}{a^2} \left (1-\cos k_z a \right) \right)
\label{dispxover}
\end{equation}
where $\xi_0 = a/\sqrt{2}$ and $k_\parallel$ is the in-plane momentum.
(More generally, $\xi_0$ is a microscopic length scale of order $1/k_F$ where
$k_F$ is the electronic Fermi momentum.)

The character of the spin fluctuations depends on whether their energy is larger or
smaller than the vertical bandwidth $t'$. In the energy range $t' \ll \omega \ll t$, the paramagnon excitation energy is mainly accounted for by the in-plane momentum $k_\parallel$, and in this sense the
spin fluctuations are effectively two-dimensional,
\begin{align} \label{dispersion2D}
\omega(\K) &\approx
t \xi^2_0 k_\parallel^2 ,
\quad {\rm for} \quad
t' \ll \omega \ll t\quad ({\rm 2d}).
\end{align}
The restriction on the energy $\omega$ implies that the in-plane momenta are
confined to the window $\Lambda_\eta \ll k_\parallel \ll \Lambda$, where $\Lambda \sim
\xi_0^{-1}$ is a momentum cutoff,
and we defined for later convenience the crossover momentum
\begin{align} \label{CrossoverMomentum}
\Lambda_\eta  = \sqrt{\eta} \Lambda.
\end{align}
For such large energies, $\omega \gg t'$,
the paramagnons are effectively dispersionless in the direction
perpendicular to the planes, such that fluctuations with all vertical momenta $k_z$ are
equivalent. On the other hand, for energies of order $t'$ and smaller, the vertical
momentum can be resolved, and the spin fluctuations have a 3d character.
For $\omega \ll t'$ a full continuum approximation is justified, i.e.
\begin{equation}
\label{dispersion3D}
\omega(\K) \approx
t \xi^2_0 \left( k_\parallel^2 + \eta k_z^2 \right),
\quad {\rm for} \quad
\omega \ll t' \quad ({\rm 3d}).
\end{equation}
In this 3d regime, the momenta are now restricted to $k_z \ll \Lambda$ and $k_\parallel \ll \Lambda_\eta$.
In both the energy ranges $\omega \ll t'$ and $t' \ll \omega \ll t$,
the limiting forms of the dispersion, (\ref{dispersion2D}) and (\ref{dispersion3D}), respectively,
are sufficient for the description of thermodynamics.
These forms are independent of microscopic details, the only
requirement being the existence of a dimensional crossover scale $\propto t'$.
In contrast, the precise properties of the crossover itself depend on details
of the band structure, i.e., Eq.~\eqref{dispxover}, and are non-universal.
As we are less interested in these non-universal details, we approximate the
dispersion by
\begin{align} \label{AnisoDispersion}
\omega(\K) = t \xi^2_0 \left\{
\begin{array}{lll}
k_\parallel^2 & {\rm for} \quad
\Lambda_\eta < k_\parallel < \Lambda & {\rm 2d}
\\[0.5em]
k_\parallel^2 + \eta k_z^2 & {\rm for} \quad
\sqrt{k_\parallel^2 + \eta k_z^2} < \Lambda_\eta
& {\rm 3d}
\end{array} \right.
\end{align}
%


We will use the dispersion (\ref{AnisoDispersion}) for the propagator of the spin fluctuations
\begin{align} \label{AnistropicSusceptibility}
\lefteqn{\chi_0^{-1}(\K,i \omega_n) =}
\\\nn& \delta_0 + \frac{|\omega_n|}{\Gamma_\K} + \xi^2_0 \left\{
\begin{array}{lll}
k_\parallel^2 & {\rm for} \quad
\Lambda_\eta < k_\parallel \ll \Lambda & {\rm 2d}
\\[0.5em]
k_\parallel^2 + \eta k_z^2 & {\rm for} \quad
\sqrt{k_\parallel^2 + \eta k_z^2} < \Lambda_\eta
& {\rm 3d}
\end{array} \right.
\end{align}
Here, $\delta_0$ represents the bare mass of the spin fluctuations,
and the dynamics of the spin fluctuations is encoded in the Landau
damping term $|\omega_n|/\Gamma_\K$ which will be discussed in the next
subsection.

Varying $\delta_0$ drives the system through the QPT, which occurs
at $\delta_0 = \delta_{0,\rm cr}$.
We define the control parameter $r$ of the QPT via
\begin{align}
r = \delta_0 - \delta_{0,\rm cr},
\end{align}
such that $r=0$ at the QCP.
To calculate observables like thermal expansion, we shall assume that
the phase transition can be tuned by changing the pressure $p$,
which is reflected in a pressure dependence of the bare mass $\delta_0(p)$.

\subsection{Landau damping}
\label{sec:damp}

The dynamics of the paramagnetic spin fluctuations is controlled by damping due to
particle--hole excitations in the metal;
this Landau damping term $|\omega_n|/\Gamma_\K$ dominates over the undamped dynamical term
$\propto\omega_n^2$ in the propagator at low energies.
The function $\Gamma_\K$ has to be obtained from an expansion of the particle--hole bubble
of the host metal, for small energies and momenta close to the ordering wavevector.
As the damping is different for ferro- and antiferromagnets, the different energy regimes
discussed above need to be distinguished.
Note that in the following we assume that the Fermi wavevector is sufficiently
large for fermions to cause damping, $2 k_F > |{\bf Q}|$.

Let us start with the dimensional crossover from 2d AFM to 3d AFM.
In both asymptotic regimes, the wavevector $\bf Q$ connects hot lines on the
Fermi surface, hence the known result for the antiferromagnet, $\Gamma_\K=$ const,
applies. As we are not interested in details of the crossover, we shall employ
\begin{align}
\Gamma_\K =
\left\{
\begin{array}{lll}
\Gamma_{2d} & {\rm for} \quad
\Lambda_\eta < k_\parallel < \Lambda,\;
& {\rm 2d\, AFM}
\\[0.5em]
\Gamma_{3d} & {\rm for} \quad
\sqrt{k_\parallel^2 + \eta k_z^2} < \Lambda_\eta,
& {\rm 3d\, AFM}.
\end{array} \right.
\end{align}

The dimensional crossover from 2d FM to 3d AFM is somewhat more complicated.
The ordering wavevector is ${\bf Q}=(0,0,Q_z)$.
In the 2d regime, i.e., for energies $\omega\gg t'$,
we assume that not only the spin fluctuation spectrum, but also
the fermionic dispersion is 2d.
Without dispersion in the vertical direction,
we obtain for $\Gamma_\K$ the standard result for the ferromagnet,
$\Gamma_\K \propto k_\parallel$.
On the other hand, in the 3d regime we again have $\Gamma_\K=$ const for the AFM.
Hence, we approximate $\Gamma_\K$ by
\begin{align} \label{damping2}
\Gamma_\K =
\left\{
\begin{array}{lll}
\Gamma_{2d}\, \xi_0 k_\parallel & {\rm for} \quad
\Lambda_\eta < k_\parallel < \Lambda,\;
& {\rm 2d\, FM}
\\[0.5em]
\Gamma_{3d} & {\rm for} \quad
\sqrt{k_\parallel^2 + \eta k_z^2} < \Lambda_\eta,
& {\rm 3d\, AFM}
\end{array} \right.
\end{align}
An explicit evaluation of the Landau damping for FM fluctuations in the presence of an anisotropic
Fermi surface is given in App.~\ref{appen:lindhard}. The result given there, Eq.~\eqref{lind_res}, reduces to
the limiting form of \eqref{damping2} in the 2d FM regime.

In principle, the damping coefficients in the two different regimes, $\Gamma_{2d}$ and
$\Gamma_{3d}$, differ from each other. However, as they  only determine the overall energy
scale, we will set them, for simplicity, equal from now on, $\Gamma = \Gamma_{2d} =
\Gamma_{3d}$ (see also the discussion in Sec.~\ref{sec:univ} below).

\subsection{Correlation length}
\label{sec:corrlen}

The quartic interaction $u_0$ among spin fluctuations in (\ref{HertzModel}) will modify its
bare susceptibility (\ref{AnistropicSusceptibility}). For our model, this modification is
captured by an effective correlation length $\xi$,
\begin{align} \label{AnistropicSusceptibilityRen}
\lefteqn{\chi^{-1}(\K,i \omega_n) =}
\\\nn& \xi^{-2} + \left\{
\begin{array}{lll}
|\omega_n| k_\parallel^{2-z}+ k_\parallel^2 & {\rm for} \quad
\Lambda_\eta < k_\parallel < \Lambda & {\rm 2d}
\\[0.5em]
|\omega_n|  + k_\parallel^2 + \eta k_z^2 & {\rm for} \quad
\sqrt{k_\parallel^2 + \eta k_z^2} < \Lambda_\eta
& {\rm 3d}
\end{array} \right.
\end{align}
where in the 2d regime the dynamical exponent is either $z=2$ for AFM or $z=3$ for FM
fluctuations. From now on, we will employ dimensionless units, i.e., we set effectively the
length scale $\xi_0 = 1$ and the energy scale $\Gamma = 1$. In addition, we use a unit
volume $V=1$.

For $d+z>4$, the correlation length $\xi$ can be obtained from a self-consistent perturbation
theory in $u_0$:
\begin{align}
\xi^{-2} = \delta_0 + \frac{N+2}{6} u_0 T \sum_{\K \omega_n} 
\chi(\K,i \omega_n).
\end{align}
Only in a regime of 2d antiferromagnetic fluctuations,
this formula misses logarithmic corrections to the correlation length.
In order to capture these, we will later have to apply the
renormalization group (RG).

Substituting the sum over Matsubara frequencies by an integral over the real axis, we obtain
\begin{align} \label{CorrelationLength}
\xi^{-2} =&\, \delta_0 + \frac{N+2}{6} u \int_0^{\Lambda_\omega} \frac{d\omega}{\pi} \coth\frac{\omega}{2 T}
\nn\\&
\left[
\int_{\Lambda_\eta}^\Lambda \frac{dk k}{2\pi}
\frac{\omega k^{2-z}}{\left(\xi^{-2} + k^2\right)^2+ \left( \omega k^{2-z}\right)^2}
\right.
\\\nn&
\left.
+ \frac{\pi}{\Lambda_\eta} \int_0^{\Lambda_\eta} \frac{dk k^2}{2\pi^2}
\frac{\omega}{\left(\xi^{-2} + k^2\right)^2+\omega^2}
\right]
\end{align}
where $\Lambda_\omega$ is an additional energy cutoff, and we introduced the
two-dimensional quartic coupling $u = u_0\,\Lambda/\pi$. The factor $\Lambda/\pi$
originates from the dummy momentum integration over the $z$ component in the 2d regime.
The contribution in the second (third) line is attributed to the spin fluctuations of
effectively two (three)-dimensional character. Correspondingly, we define a 2d (3d)
regime in the phase diagram plane where the second (third) line in
Eq.~\eqref{CorrelationLength} dominates the correlation length.
We obtain the following criterion:
\begin{align} \label{Regimes}
\begin{array}{lc}
T > \Lambda^{2(z-1)}_\eta \quad {\rm or}\quad r > \Lambda^{2 (z-1)}_\eta & {2d\; \rm regime}
\\[0.5em]
T < \Lambda_\eta^{2 (z-1)} \quad {\rm and}\quad r < \Lambda^{2 (z-1)}_\eta  & {3d\; \rm regime}
\end{array}
\end{align}
where $z=2$ for 2d AFM and $z=3$ for 2d FM spin fluctuations.


\subsection{Thermodynamics}
\label{subsec:Thermodynamics}

From the susceptibility (\ref{AnistropicSusceptibilityRen}), we can obtain the free
energy of the critical spin fluctuations
\begin{align}
F_{\rm cr} =&\, \frac{N T}{2} \sum_{\K,\omega_n} \log \chi^{-1}(\K,i \omega_n).
\end{align}
It will be convenient to absorb a factor $\Lambda/\pi$ in the units of the free energy,
$F\pi/\Lambda \to F$. Doing so, the free energy in our approximation takes the form
\begin{align} \label{CriticalFreeEnergy}
F_{\rm cr} =&\, -\frac{N}{2}
\int_0^{\Lambda_\omega} \frac{d\omega}{\pi} \coth\frac{\omega}{2 T}
\left[
\int_{\Lambda_\eta}^\Lambda \frac{dk k}{2\pi} \arctan \frac{\omega k^{2-z}}{\xi^{-2} + k^2}
\right.
\nn\\&
\left.
+ \frac{\pi}{\Lambda_\eta}
\int_0^{\Lambda_\eta} \frac{dk k^2}{2\pi^2}
\arctan\frac{\omega}{\xi^{-2} + k^2}
\right].
\end{align}
The integral in the first (second) line originates from the 2d (3d) spin fluctuations.

From the free energy, we can compute
thermodynamic properties. We will consider the specific heat, thermal expansion,
Gr\"uneisen parameter and the compressibility.
The specific-heat coefficient $\gamma$ is defined as
\begin{align}
\gamma = - \frac{\partial^2 F}{\partial T^2}.
\end{align}
The thermal expansion $\alpha$ measures the change in volume as the
temperature is changed,
\begin{equation}
\alpha = \left. \frac{1}{V}\frac{\partial V}{\partial T} \right|_{p}
= \frac{1}{V} \frac{\partial^2 F}{\partial p\, \partial T}
= - \left.\frac{1}{V} \frac{\partial S}{\partial p}\right|_{T}\,.
\end{equation}
Using a Maxwell equation, we have re-written the thermal expansion as a derivative of
entropy with respect to pressure. In principle, all parameters of the model
(\ref{HertzModel}) might be pressure dependent. However, it has been argued\cite{Zhu03}
that the most important contribution comes from the pressure dependence of the parameter
multiplying the most relevant operator in the model. Close to the quantum critical point,
this is the control parameter $r$ of the transition. Near a pressure-tuned quantum
critical point, we can expand the control parameter around the critical pressure $p_c$,
$r \approx (p-p_c)/p_0$, where $p_0$ is an {\it a priori} unknown pressure scale. In this
case, there is a contribution to thermal expansion that measures the change of entropy
upon variations of the control parameter $r$. Choosing dimensionless units, this
contribution is given by
\begin{align}
\alpha_{\rm cr} = \frac{\partial F_{\rm cr}}{\partial T \partial r}.
\end{align}
The critical Gr\"uneisen parameter $\Gamma_{\rm cr}$ is the ratio of critical thermal expansion
and specific heat,
\begin{equation}
\Gamma_{\rm cr} = \frac{\alpha_{\rm cr}}{T \gamma_{\rm cr}}\,.
\end{equation}
The compressibility $\kappa$ measures the change in volume as the pressure is
changed with temperature held fixed,
\begin{equation}
\kappa = \left. -\frac{1}{V}\frac{\partial V}{\partial p} \right|_{T}
= - \frac{1}{V} \frac{\partial^2 F}{\partial p^2}\,.
\end{equation}
The quantum critical fluctuations contribute to $\kappa$ an additive term,
in the following denoted by $\kappa_{\rm cr}$.
The pressure dependence of the control parameter $r$ results
in the following contribution to $\kappa_{\rm cr}$:
\begin{align} \label{Kappa}
\kappa_{\rm cr} &=  -\frac{\partial^2 F_{\rm cr}}{\partial r^2}
\end{align}
in dimensionless units.

\subsection{Universality?}
\label{sec:univ}

Before presenting actual results, it is worth asking how ``universal'' we can expect them
to be. This question has various aspects:
(i) Are the crossover functions (for a specific observable) universal in the sense
that they do not depend on microscopic details -- at least in a certain well-defined
limit?
(ii) Do all observables display the same crossover scale(s)?

For aspect (i) the answer is that full universality does {\em not} exist, as the quantum
phase transitions under consideration are at or above their upper critical dimension.
Therefore, even if the ultraviolet cut-off $\Lambda$ and the crossover scale
$\Lambda_\eta$ are well separated, i.e. $\eta\ll 1$, the bare values of $\Lambda$ and the
paramagnon interaction $u$ will influence the crossover functions.
In particular, the interaction $u$ is at the origin of the classical dimensional
crossover lines denoted as $\Tcl$ in Figs.~\ref{fig:pd-AFAF} and \ref{fig:pd-FMAF}.

Moreover, as discussed at length in Sec.~\ref{sec:bare},
the dimensional crossover itself is determined by microscopic details, i.e. the precise crossover form of the
bare susceptibility. At this point, possibly existing magnetic inter-layer
frustration enters.\cite{ormv}
Similarly, the reference energy scales of the 2d and 3d regimes,
$\Gamma_{2d}$ and $\Gamma_{3d}$, which we have taken to be equal for
simplicity, may differ (by a factor of order unity),
which leads to a shift of the two asymptotic regimes on the
temperature axis w.r.t. each other.

For some of the observables listed above, additional care has to be taken
regarding the dependence on external pressure. As the pressure dependence
of all microscopic parameters is smooth, singular contributions to thermodynamics
usually arise only through the pressure dependence of $r$ in the vicinity of the QCP.
However, in our model, the 2d regime possesses another relevant parameter,
namely the anisotropy $\eta$.
If the anisotropy also depends on pressure, it will give important additional
contributions to thermal expansion and compressibility in the 2d regime.
In dimensional units, these contributions are represented by
\begin{align} \label{AnisoContribution}
\alpha^\eta_{\rm cr} = 
\frac{\partial F_{\rm cr}}{\partial T \partial \eta},\qquad
\kappa^\eta_{\rm cr} = - 
\frac{\partial^2 F_{\rm cr}}{\partial \eta^2}.
\end{align}
They can be best estimated by considering the free energy $F$ with a momentum dependence
for the spin-fluctuations given by Eq.~(\ref{dispxover}), instead of the simplified
expression for $F$, (\ref{CriticalFreeEnergy}). In the 2d regime, we obtain that the
pressure dependence of the anisotropy yields contributions proportional to the ones
deriving from a pressure-dependent control parameter, $\alpha^\eta_{\rm cr} \propto
\alpha_{\rm cr}$ and $\kappa^\eta_{\rm cr} \propto \kappa_{\rm cr}$. In the 3d regime, on
the other hand, $\alpha^\eta_{\rm cr}$ and $\kappa^\eta_{\rm cr}$ are suppressed by
additional powers of momenta compared to $\alpha_{\rm cr}$ and $\kappa_{\rm cr}$,
respectively, and are therefore negligible. Having this in mind, we omit a further
discussion of these terms in the following sections.

Given all these caveats, our calculations to be presented below are nevertheless valuable,
for various reasons:
(a) they illustrate the general structure of the phase diagram with the quantum and classical dimensional crossover lines,
(b) they show the existence of the interesting intermediate regime where 2d quantum fluctuations coexist with 3d classical fluctuations,
(c) they show which observables are especially sensitive to the dimensional crossover and how ``broad'' or ``narrow'' the crossover signatures are expected to be, and
(d) they show where logarithmic corrections can dominate power laws.

Let us briefly comment on aspect (ii), namely whether all observables display the same
crossover scale. For thermodynamics, there is a single relevant momentum crossover scale
$\Lambda_\eta$, at least in our simple model without inter-layer frustration. The single
momentum scale however translates into various temperature scales, and depending on the
thermodynamic quantity of interest different crossover scales might be of importance. For
the 2d -- 3d AFM crossover, there are two relevant temperature scales: the quantum
dimensional crossover temperature $T \sim \Lambda_\eta^2$, and the classical crossover
temperature $\Tcl$, determined by the paramagnon interaction $u$. Whereas the
specific-heat coefficient is insensitive to the $\Tcl$ scale, the thermal expansion and
compressibility show signatures at both scales, and their dimensional crossover appears to be broad due to the existence of the intermediate region, $\Lambda^2_\eta < T < \Tcl$. For the 2d FM -- 3d AFM crossover, we find even three temperature scales arising from the single momentum scale $\Lambda_\eta$:
a quantum dimensional crossover temperature $T \sim \Lambda_\eta^4$, a
temperature scale $T \sim \Lambda_\eta^3$ dividing 2d non-critical from 2d quantum
critical behavior, and a classical dimensional crossover scale $T \sim \Tcl$. These
multiple crossover temperatures result in rich thermodynamics and render the extraction
of power laws especially difficult.

Transport properties -- which are not subject of this paper -- can be expected
to display even more complicated crossover behavior. In addition to the momentum scale
$\Lambda_\eta$, there are characteristic length scales for transport scattering
processes. For coupled chains, it has been argued\cite{rosch-chains} that the interplay of those length and time
scales can render transport fully three-dimensional even in a regime where the
thermodynamic behavior is 1d. Hence, thermodynamic and transport crossovers do not have to coincide in general.



\section{Dimensional crossover: 2d antiferromagnet to 3d antiferromagnet}
\label{sec:2af3af}

For materials consisting of antiferromagnetic planes,
which itself are weakly coupled in the third direction,
a scenario of a crossover from 2d antiferromagnetism to 3d antiferromagnetism is plausible. As mentioned in the introduction, such a scenario might be realized in the heavy-fermion metal \CeAu. In the following, we present an analysis of this crossover within the framework of the LGW model whose applicability to \CeAu, however, has been questioned.\cite{si,coleman} 

\subsection{Renormalization group}

The Hertz model (\ref{HertzModel}) for AFM fluctuations in 2d is at its upper critical
dimension, $d+z = d^+_c = 4$.
Consequently, there are important logarithmic corrections in
perturbation theory that have to be summed, e.g., with the help of the renormalization
group (RG). The RG equations for the running mass $\delta(b)$ and quartic coupling $u(b)$
in the 2d AFM regime read:\cite{hertz,millis,hvl}
\begin{subequations}
\label{RGEquations}
\begin{align}  \label{RGEquations3}
\frac{\partial\, \delta}{\partial \log b} &= 2 \delta
-  \frac{N+2}{12 \pi^2}\,u \delta \,,
\\ \label{RGEquations4}
\frac{\partial\, u}{\partial \log b} &=
-  \frac{N+8}{12 \pi^2}\, u^2\,.
\end{align}
\end{subequations}
Here, we have introduced the RG scale $b$, and the RG flow corresponds to a reduction of
the momentum-space cutoff, $\Lambda \to \Lambda/b$.
The one-loop correction to $\delta$, i.e., the tadpole diagram,
can be expanded in a power series in $\delta$:
The constant term is finite and non-universal (i.e. cutoff-dependent),
whereas the linear term is universal.
The latter linear term is written as second term in Eq.~\eqref{RGEquations3};
the constant term will be absorbed in the initial value of $\delta$.
Hence, the flow starts at $b=1$ with the initial conditions
$\delta(b\!=\!1) = r_{2d}$ (that differs from $\delta_0$ by the non-universal Hartree shift)
and $u(b\!=\!1) = u = u_0 \Lambda/\pi$, being the effective 2d quartic coupling.
Note that the prefactors in \eqref{RGEquations} have been evaluated at $T=0$,
as their temperature dependence is subleading.

Integrating the RG equation for the quartic coupling we obtain
\begin{eqnarray} \label{QuarticCoupling-LimitingForm}
u(b) =
\frac{12 \pi^2}{N+8}\,
\frac{1}{\log \frac{b \bar{\Lambda}}{\Lambda}},
\end{eqnarray}
where we have introduced the momentum scale $\bar{\Lambda}$,
which depends on the bare quartic coupling constant $u$,
\begin{equation} \label{RenormalizedCutoff}
\bar{\Lambda} \equiv
\Lambda\,\me^{\textstyle \frac{12 \pi^2}{(N+8) u}  }\,.
\end{equation}
Using the running quartic coupling, the solution for the control parameter is easily
obtained. Substituting $\delta(b) = r(b) b^2$, we get
\begin{align} \label{Mass2DFL}
r(b) =
\frac{r_{2d}}
{\left(
\log \left[
b\; \frac{\bar{\Lambda}}{\Lambda}\right]\right)^
{\frac{N+2}{N+8}}}.
\end{align}
As we will see later, the parameter $r_{2d}$ differs from the control parameter, $r \equiv r_{3d}$, of the quantum phase transition by corrections of order $\Lambda_\eta^2$.
%

\subsection{Correlation length}

The logarithmic scale dependence of the coupling and the control parameter
has to be in calculating the correlation length.
Hence, Eq.~(\ref{CorrelationLength}) is replaced by
\begin{align} \label{CorrelationLengthAFM}
\lefteqn{\xi^{-2} = r({\Lambda/\Lambda^*})    }
\\\nn &\,  + \frac{N+2}{6}  u({\Lambda/\Lambda^*})
\int_0^{\infty} \frac{d\omega}{\pi}
\int_{\Lambda_\eta}^\Lambda \frac{dk k}{2\pi}
\frac{\omega \left(\coth\frac{\omega}{2 T} - 1\right)}{\left(\xi^{-2} + k^2\right)^2+ \omega^2}
\\\nn&
+ \frac{N+2}{6} \frac{\pi u^*}{\Lambda_\eta} \int_0^{\Lambda_{\omega}} \frac{d\omega}{\pi}
 \int_0^{\Lambda_\eta} \frac{dk k^2}{2\pi^2}
\frac{\omega \coth\frac{\omega}{2 T}}{\left(\xi^{-2} + k^2\right)^2+\omega^2}.
\end{align}
The abbreviations $\Lambda^*$ and $u^*$ are defined through
\begin{equation}
\Lambda^* ={\rm max}\{\xi^{-1},\Lambda_{\eta}\}
\label{lamst}
\end{equation}
and
\begin{align} \label{CrossoverCoupling}
u^* = u(\Lambda/\Lambda_\eta)  =  \frac{12 \pi^2}{N+8}\,
\frac{1}{\log \frac{\bar{\Lambda}}{\Lambda_\eta}},
\end{align}
the latter being the quartic coupling at the crossover scale $\Lambda_\eta$.
Eq.~\eqref{CorrelationLengthAFM} can be understood as follows:
The first line accounts for the 2d quantum fluctuations at $T=0$
which have been re-summed in $r(b)$ using the RG, see Eq.~(\ref{Mass2DFL}).
Finite-temperature corrections to this 2d result can be treated perturbatively\cite{ssbook}
and are in the second line.
The third line, finally, is attributed to the 3d fluctuations.

In the following, the limiting behavior of the correlation length is discussed in detail.

\subsubsection{Correlation length in the 2d regime, $T \gg \Lambda_\eta^2$ or $r_{2d} \gg \Lambda_\eta^2$}

In the 2d regime as defined in Eq.~(\ref{Regimes}), we can neglect the last line in the expression
(\ref{CorrelationLengthAFM}) as it yields only a small correction. We further
distinguish two sub-regimes.

{\it Fermi-liquid regime, $T \ll r_{2d}$:}
In the Fermi-liquid regime, the correlation
length is determined by the solution of the RG equation for the tuning parameter
\begin{equation}
\xi^{-2} =
\frac{r_{2d}}{\left(\log \frac{\bar{\Lambda}}{\sqrt{r_{2d}}}
\right)^{\frac{N+2}{N+8}}}.
\end{equation}
The logarithmic dependence can be understood as an incipient correction to the mean-field
value of the correlation length exponent $\nu_{\rm MF}=1/2$.

{\it Quantum critical regime, $T \gg r_{2d}$:}
The limiting behavior of the remaining integral is given by the small $\omega$ limit of
the integrand, such that
\begin{align}
\xi^{-2} =  r(\Lambda /\Lambda^*) + \frac{\pi}{2} \frac{N+2}{N+8} T \frac{\log
\frac{T}{{\Lambda^*}^2}}{\log\frac{\bar{\Lambda}}{\Lambda^*}}.
\end{align}
Hence, in the quantum critical regime, the correlation length is determined by
temperature, $\xi^{-2} \sim T$, up to logarithmic corrections. The logarithms are either
cut-off by the correlation length itself or by the dimensional crossover scale $\Lambda_\eta$.
In the limit $\xi^{-2} \gg \Lambda_\eta^2$, the correlation length is asymptotically given by
\begin{align} \label{CorrelationLength2dQC1}
\xi^{-2} = r\left(\frac{\Lambda}{\sqrt{T}}\right) + \frac{\pi}{2} \frac{N+2}{N+8}  T
\frac{\log \log \frac{\bar{\Lambda}}{\sqrt{T}}}{\log \frac{\bar{\Lambda}}{\sqrt{T}}},
\quad{\rm for}\,\,\xi^{-2} \gg \Lambda_\eta^2.
\end{align}
The $\log \log$ dependence in the numerator can be attributed to the 2d classical
fluctuations, i.e, the 2d-like Matsubara zero mode, within the quantum critical regime.
[Depending on whether the first or the second term dominates in
(\ref{CorrelationLength2dQC1}), one can further distinguish two sub-regimes within the regime
denoted as the renormalized 2d classical regime in Fig.~\ref{fig:pd-AFAF}.]
When the correlation length exceeds the crossover scale, i.e. $\xi^{-2} \ll \Lambda_\eta^2$, the
logarithms are now cut-off by $\Lambda_\eta$,
\begin{align} \label{CorrelationLength2dQC2}
\xi^{-2} = r(\Lambda/\Lambda_\eta) + \frac{\pi}{2} \frac{N+2}{N+8} T \frac{\log \frac{T}{\Lambda^2_\eta} }{\log \frac{\bar{\Lambda}}{\Lambda_\eta}},
\quad{\rm for}\quad \xi^{-2} \ll \Lambda_\eta^2.
\end{align}

The crossover at the scale $\xi^{-2} \sim \Lambda_\eta^2$ is associated with
the advocated dimensional crossover for the classical critical fluctuations,
where the Matsubara zero mode changes its character from 2d to 3d.
At criticality, $r_{2d} = 0$, this occurs at a temperature
\begin{align}
\left. \Tcl \right|_{r_{2d}=0} \sim
\Lambda^2_{\eta} \frac{\log\frac{\bar{\Lambda}}{\Lambda_\eta}}{\log \log \frac{\bar{\Lambda}}{\Lambda_\eta}}.
\end{align}
This classical dimensional crossover temperature, $\Tcl$, is logarithmically enhanced as
compared to the quantum dimensional crossover temperature, $T \sim \Lambda^2_\eta$, see
Fig.~\ref{fig:pd-AFAF}.

The classical dimensional crossover occuring at $\Tcl$ becomes non-perturbative when the line $\Tcl(r)$ enters the classical Ginzburg region. Generally, our perturbative treatment breaks down sufficiently close to the classical transition at the Ginzburg temperature $T_G$, where the classical Ginzburg parameter is of order one, $\mathcal{G} \sim \mathcal{O}(1)$. For 3d classical fluctuations, the Ginzburg parameter is given by $\mathcal{G} = U \xi$, where the classical quartic coupling is $U = \pi u^* T/\Lambda_\eta \sim T/(\Lambda_\eta \log \frac{\bar{\Lambda}}{\Lambda_\eta})$. Upon increasing temperature, the crossover line $\Tcl(r)$ approaches $T_G(r)$ and enters the Ginzburg region at a temperature $T_x$ where $T_G$ and $\Tcl$ coincide, see Fig.~\ref{fig:pd-AFAF},
\begin{align} \label{CrossingT-AFAF}
T_x \sim \Lambda^2_{\eta} \log\frac{\bar{\Lambda}}{\Lambda_\eta}.
\end{align}


\subsubsection{Correlation length in the 3d regime, $T \ll \Lambda_\eta^2$ and $r_{3d} \ll \Lambda_\eta^2$}

In the 3d regime, we can instead neglect the second line in (\ref{CorrelationLengthAFM}).
We again distinguish between two sub-regimes.

{\it Fermi-liquid regime, $T \ll r_{3d}$:}
In the Fermi-liquid regime, we can set $T=0$ in (\ref{CorrelationLengthAFM}) in order to
obtain the leading estimate for the correlation length,
\begin{align}
\label{ControlParameter3d}
\xi^{-2} = r_{3d} \equiv
\frac{r_{2d}}{\left(\log \frac{\bar{\Lambda}}{\Lambda_\eta} \right)^{\frac{N+2}{N+8}}}
+ \mathcal{C} \frac{\Lambda_\eta^2}{\log \frac{\bar{\Lambda}}{\Lambda_\eta}},
\end{align}
where $\mathcal{C}$ is a non-universal constant that depends on the chosen cut-off
structure. The control parameter $r \equiv r_{3d}$ obtains a shift from the contribution of the
parameter regime where the spin fluctuations have developed their three-dimensional
character; the consequences of this shift are discussed in
Sec.~\ref{sec:PhaseBoundaryAFM-AFM}. The temperature correction neglected in
(\ref{ControlParameter3d}) is of Fermi-liquid type $\sim T^2$.

{\it Quantum critical regime, $T\gg r_{3d}$:}
In the quantum critical regime, the temperature
corrections dominate the correlation length. Evaluating the leading behavior we obtain
\begin{align} \label{CorrelationLengthExplicit3dQC}
\xi^{-2} = r_{3d} + \left(\frac{\pi}{2}\right)^{3/2}\,\zeta(3/2)
\frac{N+2}{N+8}
\frac{T^{3/2}}{\Lambda_\eta \log \frac{\bar{\Lambda}}{\Lambda_\eta}},
\end{align}
where we used the explicit expression for $u^*$, Eq.~(\ref{CrossoverCoupling}). The
quantum critical regime can again be further subdivided into regimes where either of the
above two terms dominates the correlation length.

\subsubsection{Phase boundary}
\label{sec:PhaseBoundaryAFM-AFM}

The position of the phase boundary, $T_c(r)$, in the $(T,r)$ plane can be
estimated by setting the correlation length to infinity. In the quantum 2d regime for temperatures lower than the crossing temperature $T < T_x$, Eq.~(\ref{CrossingT-AFAF}), the classical dimensional crossover from 2d to 3d is perturbative as it is located outside the classical Ginzburg region. Here, we can use expression (\ref{CorrelationLength2dQC2}) for the correlation length to obtain
\begin{align} \label{PhaseBoundary2d}
r_{2d} = -  \frac{\pi}{2} \frac{N+2}{N+8}
\frac{T_c \log \frac{T_c}{\Lambda^2_\eta} }{\left(\log
\frac{\bar{\Lambda}}{\Lambda_\eta}\right)^{1-\frac{N+2}{N+8}}}.
\end{align}
For larger $T > T_x$, the logarithmic corrections to the critical temperature $T_c$ will differ from Eq.~(\ref{PhaseBoundary2d}) due to the non-perturbative character of the classical dimensional crossover. Neglecting the logarithmic corrections, the phase boundary varies linearly with temperature, $T_c \propto r_{2d}$, in the 2d regime.

\begin{figure}
\includegraphics[width=0.47\textwidth]{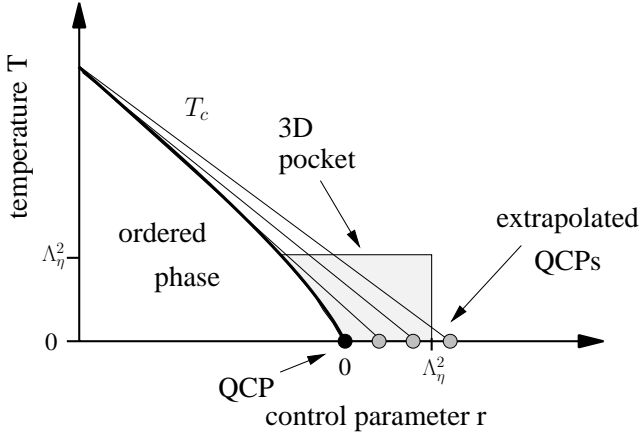}
\caption{
Locating the QCP based on experimental data requires an extrapolation of the finite-$T$
phase boundary to $T=0$.
Such an extrapolation procedure is prone to a systematic error if a dimensional crossover
at lower temperatures takes place. The extrapolation of the phase boundary (thick line) from the 2d regime
towards $T=0$ yields estimates of the position of the QCP that are
shifted towards the non-ordered phase by an amount $\Delta r$.
The different thin lines and the associated extrapolated QCPs illustrate that the extrapolation
itself is ambiguous as the 2d phase boundary is not strictly linear in $T$ but has logarithmic
corrections.
For the 2d -- 3d AFM crossover, both the size of the 3d pocket
and the shift $\Delta r$, Eq.~(\ref{ErrorExtrapolation}), are of order $\Lambda_\eta^2$.
For the 2d FM -- 3d AFM crossover of Sec.~\ref{sec:2f3af}, the size of the
3d pocket is instead given by $\Lambda^4_\eta$, while the shift $\Delta r \propto
\Lambda^2_\eta$, Eq.~(\ref{ErrorExtrapolation2}). Consequently, the extrapolated QCP is
outside the 3d pocket in this case.
}
\label{fig:asd}
\end{figure}

In the 3d regime, on the other hand, we use expression
(\ref{CorrelationLengthExplicit3dQC}) and get
\begin{align}
\label{PhaseBoundary3d}
r_{3d} = - \left(\frac{\pi}{2}\right)^{3/2}\,\zeta(3/2)
\frac{N+2}{N+8}
\frac{T_c^{3/2}}{\Lambda_\eta \log \frac{\bar{\Lambda}}{\Lambda_\eta}}.
\end{align}
The different behavior of the phase boundary in the 2d and 3d regimes,
\eqref{PhaseBoundary2d} and \eqref{PhaseBoundary3d}, implies that an extrapolation of the
phase boundary from the 2d regime leads to an incorrect position of the quantum critical
point. The 3d spin-fluctuations shift the quantum critical point slightly towards the
ordered phase. The quantum critical point is not at $r_{2d} = 0$, but rather at
$r_{3d} = 0$, leading to a difference between the extrapolated and the actual
position of the QCP,
\begin{align} \label{ErrorExtrapolation}
\Delta r = \left.r_{3d}\right|_{r_{2d} = 0}
\sim
\frac{\Lambda_\eta^2}{\log\frac{\bar{\Lambda}}{\Lambda_\eta}}.
\end{align}
The position of the extrapolated QCP, $\Delta r$, see Fig.~\ref{fig:asd},
is, however, located close to or even within the 3d pocket of the phase diagram.

\subsection{Thermodynamics}
\label{sec:ThermodynamicsAFM-AFM}

We turn to a discussion of the thermodynamic quantities specified in
Sec.~\ref{subsec:Thermodynamics}.
They can be obtained with the help of expression (\ref{CriticalFreeEnergy})
for the free energy, with the correlation length given by the self-consistent
Eq.~(\ref{CorrelationLengthAFM}). A numerical solution for the specific heat, thermal
expansion, and compressibility in the quantum critical regime is presented in
Fig.~\ref{fig:2af3af}. A detailed discussion of their asymptotic behavior is given below.

\begin{figure}
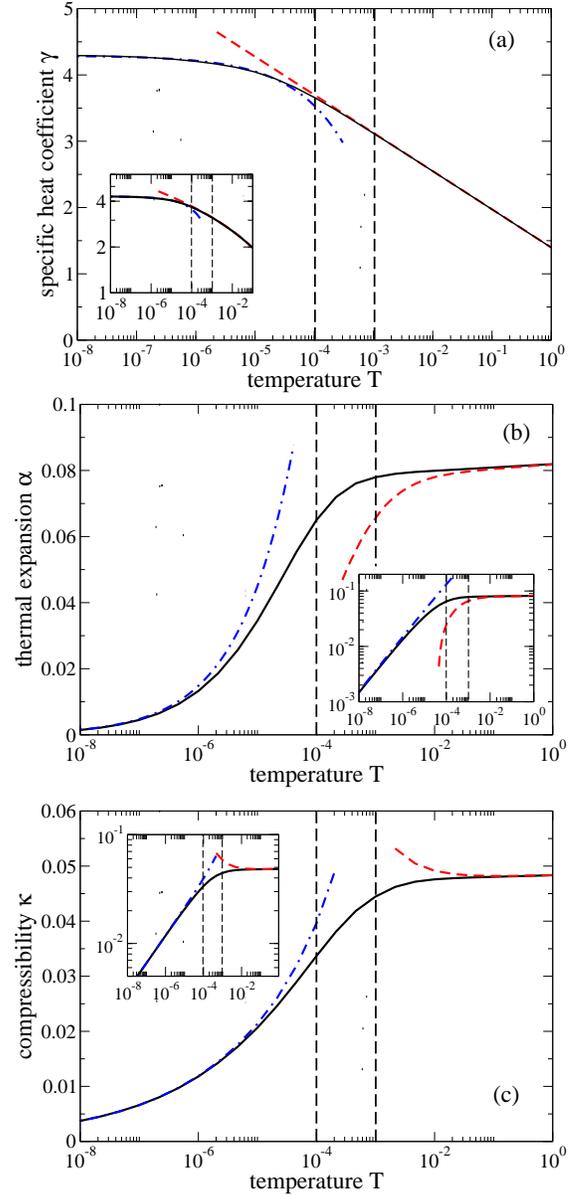

\hspace{0.5em}
\includegraphics[width=0.393\textwidth]{fig5a.eps}
\vspace{0.5em}
\includegraphics[width=0.41\textwidth]{fig5b.eps}
\vspace{0.5em}
\includegraphics[width=0.41\textwidth]{fig5c.eps}
\caption{ (color online)
Crossover behavior of thermodynamics for $r=0$, i.e., in the
quantum critical regime of the 2d AFM -- 3d AFM crossover scenario. The vertical dashed
line at higher $T \sim \Tcl$ indicates the classical and the one
at lower $T \sim \Lambda_\eta^2$ the quantum dimensional crossover temperature.
The insets show the same data on a double-logarithmic scale.
The chosen parameters are $\Lambda_\omega = \Lambda = 20$, $\Lambda_\eta = 0.01$, $u=1$ and $N=3$.
Panel (a): Specific-heat coefficient $\gamma$.
Also shown are the asymptotic behaviors given by
Eq.~(\ref{SpecificHeatQCR-d=z}) (red/dashed) for the high-temperature 2d regime and
Eq.~(\ref{SpecificHeatQCR-d>z}) (blue/dash-dot) for the low-temperature 3d regime.
Panel (b): Thermal expansion $\alpha$.
The asymptotic behaviors are given by Eq.~(\ref{ThermExpQCR-d=z=2}) (red/dashed)
and by Eq.~(\ref{ThermExpQCR-d>2}) (blue/dash-dot).
Panel (c): Compressibility $\kappa$.
The asymptotics are in Eq.~(\ref{KompQCR-AFAF-d=2}) (red/dashed) and
Eq.~(\ref{Compressibility3dQCAFM-AFM}) (blue/dash-dot).
Note the very weak increase with $T$ in
the high-temperature 2d regime.}
\label{fig:2af3af}
\end{figure}

\subsubsection{Thermodynamics in the 2d regime, $T \gg \Lambda_\eta^2$ or $r_{2d} \gg \Lambda_\eta^2$}

Within the 2d regime, the thermodynamics is dominated by the first line in the expression for
the free energy Eq.~(\ref{CriticalFreeEnergy}). We again distinguish between two
sub-regimes.

{\it Fermi-liquid regime, $T \ll r_{2d}$:}
The evaluation of the specific heat and thermal expansion is straightforward
\begin{align} 
\gamma_{\rm cr}
=  \frac{N}{6} \,
 \log \frac{\Lambda}{\sqrt{r_{2d}}},
\qquad
\alpha_{\rm cr} = \frac{N}{12} \frac{T}{r_{2d}}.
\end{align}
%
The resulting Gr\"uneisen parameter reads
\begin{align} \label{Grueneisen2dFL}
\Gamma_{\rm cr} = \frac{1}{2} \frac{1}{r_{2d} \log \frac{\Lambda}{\sqrt{r_{2d}}}}.
\end{align}
%

The evaluation of the compressibility is more involved. Here, we have to take into
account the effective momentum dependence of the correlation length, $\xi_k^{-2} =
r(\Lambda/k)$, see Eq.~(\ref{Mass2DFL}), in the expression for the free energy
(\ref{CriticalFreeEnergy}) that arises after RG improvement of perturbation theory.
We obtain
\begin{align}
\kappa_{\rm cr} &= \frac{N}{4\pi^2} \int_{\Lambda^*} \frac{dk}{k} \left(\frac{\partial
r(\Lambda/k)}{\partial r_{2d}}\right)^2.
\end{align}
The asymptotic behavior originates from the lower limit of the momentum integral that
has to be cutoff either by the crossover scale or the correlation length itself,
$\Lambda^* = {\rm max}\{\xi^{-1},\Lambda_{\eta}\}$.
In the 2d regime we thus obtain for
the asymptotic behavior:\cite{Larkin69,Wegner73}
\begin{align} \label{KompFL-2D-expl}
\kappa_{\rm cr} =\frac{N}{4\pi^2}\frac{1}{1-2\frac{N+2}{N+8}}
\left(
\log \frac{\bar{\Lambda}}{\sqrt{r_{2d}}} \right)^
{1-2\frac{N+2}{N+8}}\,.
\end{align}
%
Note that, for $N <4$, Eq.~(\ref{KompFL-2D-expl}) predicts a correction to the
compressibility that diverges upon approaching the quantum critical point, $r_{2d} \to
0$. This divergence is only cutoff upon entering the 3d regime.

A diverging electronic compressibility has interesting consequences.
In particular, a coupling of the electronic system to lattice degrees of freedom can
render the coupled system unstable, resulting in a first-order transition
driven by quantum critical AFM fluctuations.\cite{Anfuso08}
Such a fluctuation-driven first-order transition may occur in the present case
for a sufficiently large 2d regime.

{\it Quantum critical regime, $T \gg r_{2d}$:}
In the 2d quantum critical regime,
the specific heat depends logarithmically on temperature,
\begin{eqnarray} \label{SpecificHeatQCR-d=z}
\gamma_{\rm cr} =  \frac{N}{6} \log \frac{\Lambda}{\sqrt{T}} \,.
\end{eqnarray}
For the thermal expansion we obtain
\begin{align} \label{ThermExpQCR-d=z=2}
\alpha_{\rm cr} =\frac{N}{8 \pi} \frac{\partial \xi^{-2}}{\partial r_{2d}}
\left( \log \frac{T}{\xi^{-2}+\Lambda_\eta^2} + {\rm const} \right).
\end{align}
This leads to slightly different asymptotic behavior within the 2d and 3d classical regime, see Fig.~\ref{fig:pd-AFAF},
\begin{align}
\alpha_{\rm cr}
\sim
\left\{\begin{array}{ll}
\left(\log \frac{\bar{\Lambda}}{\sqrt{T}}\right)^
{-\frac{N+2}{N+8}} \log \log \frac{\bar{\Lambda}}{\sqrt{T}} & \quad{\rm if}\qquad T\gg \Tcl\\[0.5em]
\left(\log \frac{\bar{\Lambda}}{\Lambda_\eta}\right)^
{-\frac{N+2}{N+8}} \log \frac{T}{\Lambda_\eta^2} & \quad{\rm if}\qquad T\ll \Tcl
\end{array}
\right. .
\end{align}
For $T \gg \Tcl$, we obtain for the Gr\"uneisen parameter the temperature dependence
\begin{align}
\Gamma_{\rm cr} \sim
\frac{\log \log \frac{\bar{\Lambda}}{\sqrt{T}} }
{T \log \frac{\Lambda}{\sqrt{T}}\left(
\log \frac{\bar{\Lambda}}{\sqrt{T}}\right)^
{\frac{N+2}{N+8}}} \,.
\end{align}

The critical part of the compressibility
%
is dominated by the Matsubara
zero mode. Its form also differs in the classical 2d and 3d regime,
\begin{align} \label{KompQCR-AFAF-d=2}
\kappa_{\rm cr} = \left(\frac{\partial \xi^{-2}}{\partial r_{2d}} \right)^2 \times
\left\{
\begin{array}{ll}
\frac{N}{8\pi} \frac{T}{\xi^{-2}}
& {\rm if}\quad T \gg \Tcl 
\\[0.5em]
\frac{N}{16} \frac{T}{\Lambda_\eta \xi^{-1}}
& {\rm if}\quad T \ll \Tcl 
\end{array}
\right.
\end{align}
Using the expressions for the correlation length (\ref{CorrelationLength2dQC1}) and
(\ref{CorrelationLength2dQC2}), we obtain the asymptotic behavior at criticality $r_{2d}
= 0$,
\begin{align}
\kappa_{\rm cr} \sim
\left\{
\begin{array}{ll}
\left(\log \frac{\bar{\Lambda}}{\sqrt{T}} \right)^{-2\frac{N+2}{N+8}}
\frac{\log \frac{\bar{\Lambda}}{\sqrt{T}}}{\log \log \frac{\bar{\Lambda}}{\sqrt{T}}}
& {\rm if}\quad T \gg \Tcl 
\\[1em]
 \frac{\sqrt{T}}{\Lambda_\eta} \left(\log \frac{\bar{\Lambda}}{\Lambda_\eta} \right)^{-2\frac{N+2}{N+8}} \sqrt{ \frac{\log \frac{\bar{\Lambda}}{\Lambda_\eta}}{\log \frac{T}{\Lambda_\eta^2}} }
& {\rm if}\quad T \ll \Tcl 
\end{array}
\right.
\end{align}
The asymptotics of thermal expansion and compressibility in the logarithmically small
intermediate regime $\Lambda_\eta^2 < T < \Tcl$ are not shown in Fig.~\ref{fig:2af3af}.

\subsubsection{Thermodynamics in the 3d regime, $T \ll \Lambda_\eta^2$ and $r_{3d} \ll \Lambda_\eta^2$}

In the 3d regime, the critical contributions to thermodynamics are coming from the second
line in Eq.~(\ref{CriticalFreeEnergy}).

{\it Fermi-liquid regime, $T \ll r_{3d}$:}
Here, the specific heat has the form
\begin{align}\label{SpecificHeatFL2d}
\gamma &= \frac{N}{6} \log\frac{\Lambda}{\Lambda_\eta} + \gamma_{\rm cr},\quad
\gamma_{\rm cr} =
- \frac{N \pi}{12} \frac{\sqrt{r_{3d}}}{\Lambda_\eta}.
\end{align}
The critical part, $\gamma_{\rm cr}$, that depends on the control parameter, $r_{\rm
3d}$, leads here only to a small correction to the background contribution that
originates from the 2d fluctuations. The thermal expansion reads
\begin{align}
\alpha_{\rm cr} &= \frac{N\pi}{24} \frac{T}{\Lambda_\eta\sqrt{r_{3d}}}.
\end{align}
We omit here and in the following multiplicative factors that are powers of
$\partial \xi^{-2}/\partial r_{2d}$, that includes the logarithmic
normalization of the 2d control parameter, see (\ref{ControlParameter3d}).
In the Fermi-liquid regime the critical Gr\"uneisen ratio,
$\Gamma_{\rm cr} = \alpha_{\rm cr}/(T \gamma_{\rm cr})$, reads
%
\begin{align} \label{Grueneisen-AFAF3d}
\Gamma_{\rm cr} &= -\frac{1}{2} \frac{1}{r_{3d}}.
\end{align}
As explained in Ref.~\onlinecite{Zhu03}, in the Fermi-liquid regime scaling predicts a
universal critical Gr\"uneisen ratio in the sense that the proportionality factor in the
relation $\Gamma_{\rm cr} \propto 1/r$ is just determined by critical exponents. The prefactor $-1/2$ in Eq.~(\ref{Grueneisen-AFAF3d}) is in agreement
with this scaling prediction.
For the compressibility we get
\begin{align} \label{Compressibility3dFLAFM-AFM}
\kappa_{\rm cr} = -\frac{N}{8 \pi} \frac{\sqrt{r_{3d}}}{\Lambda_\eta}.
\end{align}

{\it Quantum critical regime, $T \gg r_{3d}$:}
Again, the critical part of the specific heat is
only sub-leading (\ref{SpecificHeatFL2d}) now with
\begin{align}\label{SpecificHeatQCR-d>z}
\gamma _{\rm cr} =
- \frac{15 \zeta(5/2) N}{\sqrt{2 \pi} 32} \frac{\sqrt{T}}{\Lambda_\eta} .
\end{align}
The thermal expansion is given by
\begin{align}
\label{ThermExpQCR-d>2}
\alpha_{\rm cr} =
N  \frac{3 \zeta(3/2) }{\sqrt{2 \pi} 16} \frac{\sqrt{T}}{\Lambda_\eta}.
\end{align}
The thermal expansion behaves as $\sqrt{T}$ in the 3d regime. The sub-leading correction
to (\ref{ThermExpQCR-d>2}) vanishes as $T^{3/4}$, with a two-fold origin.
First, there is a contribution due to the next-to-leading term in the expansion of the second
derivative of the free energy $\partial^2 F_{\rm cr}/\partial T\partial \xi^{-2}$ that is
of order $\xi^{-1}/\Lambda_\eta$.
Second, and more importantly, there is a $T$-dependent
correction originating from the derivative $\partial \xi^{-2}/\partial r_{\rm 3d}$ that
contributes to $\alpha$ a term of order $T^{3/2}/\xi^{-1}\Lambda_\eta^2$.
Both contributions originate from the $T$ dependence of the correlation length,
induced by the bosonic interaction $u$.\cite{NoteOnAlpha}
The critical Gr\"uneisen parameter deriving from (\ref{SpecificHeatQCR-d>z}) and
(\ref{ThermExpQCR-d>2}) obeys scaling, $\Gamma_{\rm cr} \sim 1/T$.

The compressibility is dominated by the Matsubara zero mode,
\begin{align} \label{Compressibility3dQCAFM-AFM}
\kappa_{\rm cr} = \frac{N}{16} \frac{T \xi}{\Lambda_\eta}
\sim
\left\{\begin{array}{ll}
\sqrt{\frac{\log \frac{\bar{\Lambda}}{\Lambda_\eta}}{\Lambda_\eta}}
\, T^{1/4}
& {\rm if}\qquad
r_{3d} \ll \frac{T^{3/2}}{\Lambda_\eta \log \frac{\bar{\Lambda}}{\Lambda_\eta}}
\\
\frac{T}{\sqrt{r_{3d}} \Lambda_\eta}
&
 {\rm if}\qquad
r_{3d} \gg \frac{T^{3/2}}{\Lambda_\eta \log \frac{\bar{\Lambda}}{\Lambda_\eta}}
\end{array}
\right.
\end{align}
where we used the expression (\ref{CorrelationLengthExplicit3dQC}) for the correlation
length. There exist two sub-regimes where the compressibility varies either linearly with
$T$ or, very close to the quantum critical point, as $T^{1/4}$
(Ref.~\onlinecite{FischerRosch}).
Note that the critical contribution to the compressibility, $\kappa_{\rm cr}$, changes
sign upon crossing over from the Fermi liquid (\ref{Compressibility3dFLAFM-AFM}) to the
quantum critical regime (\ref{Compressibility3dQCAFM-AFM}).



\section{Dimensional crossover: 2d ferromagnet to 3d antiferromagnet}
\label{sec:2f3af}

We now turn to the dimensional crossover from a 2d ferromagnet to
a 3d antiferromagnet, describing the situation of weakly AFM coupled
ferromagnetic planes, where the 3d ordered state corresponds to so-called A-type antiferromagnetism.

The key difference to the crossover in Sec.~\ref{sec:2af3af}
is related to the form of Landau damping and, as a
consequence, the different dynamical exponents $z=3$ and $z=2$ in the 2d and 3d regimes,
respectively.
The combination of low spatial dimensionality, $d=2$, and large dynamical
exponent, $z=3$, results in strong thermodynamic contributions of the 2d FM
spin fluctuations. Even below the temperature $T \lesssim \Lambda_\eta^3$,
thermally excited 2d fluctuations lead to a large non-critical Fermi liquid background
that dominates over the critical 3d AFM fluctuations in a wide parameter range, see
Fig.~\ref{fig:pd-FMAF}.
This leads to a peculiar situation at criticality, $r=0$:
the 2d and 3d quantum critical regions are separated by a temperature regime
$\Lambda_\eta^4 < T <\Lambda_\eta^3$ where non-critical Fermi liquid behavior prevails.

Other differences between the present FM -- AFM crossover and the AFM -- AFM crossover
of Sec.~\ref{sec:2af3af} are:
For the FM -- AFM crossover, the uniform susceptibility changes from critical to non-critical
behavior, see Sec.~\ref{Sec:susceptibility}.
Finally, the effective dimensionality is always larger than the upper critical dimension,
$d+z >4$.
Therefore, we can use directly Eqs. (\ref{CorrelationLength}) and (\ref {CriticalFreeEnergy}) for the
correlation length and free energy, respectively, without the need of a RG improvement.
This simplifies the analysis considerably.

\subsection{Correlation length}

In the analysis of the asymptotic behavior of the correlation length, we again
distinguish between the 2d and the 3d regime, where $\xi$ is dominated by the first and
second integral in (\ref{CorrelationLength}), respectively.

\subsubsection{Correlation length in the 2d regime, $T \gg \Lambda_\eta^4$ or $r_{2d} \gg \Lambda_\eta^4$}

{\it Fermi-liquid regime, $T\ll r_{2d}^{3/2}$ and $r_{2d} \gg \Lambda_\eta^2$:}
Evaluating the correlation length in the 2d Fermi-liquid regime we obtain
\begin{align}
\xi^{-2} = r_{2d} + \frac{\pi(N+2)}{144} u \frac{T^2}{r^{3/2}_{2d}}
\end{align}
where $r_{2d}$ differs from the bare mass $\delta_0$ by a constant cutoff-dependent shift.
Here, the temperature-dependent part is of the Fermi-liquid form and subleading.

{\it Quantum critical regime, $T\gg r^{3/2}$ and $T \gg \Lambda_\eta^3$:}
In the quantum critical regime, on the other hand, temperature dominates the correlation length
\begin{align} \label{CorrelationLengthExplicit2dQC-AFMFM}
\xi^{-2} =
r_{2d} + \frac{N+2}{24 \pi} u T \log \frac{1}{{\Lambda^\ast}^2 T^{-2/3}}
\end{align}
with $\Lambda^* ={\rm max}\{\xi^{-1},\Lambda_{\eta}\}$ as above, Eq.~\eqref{lamst}.
The 2d quantum critical regime can be subdivided into three regimes depending on how $\xi^{-2}$
compares with $\Lambda_\eta^2$, and whether the first or the temperature-dependent second
term dominates. At $\xi^{-2} \sim \Lambda^2_\eta$, a dimensional crossover associated
with the classical critical fluctuations takes place. At criticality, $r \approx
r_{2d}=0$, this crossover occurs at a temperature of order
\begin{align}
\left.\Tcl\right|_{r=0} \sim \frac{\Lambda_\eta^2}{u \log \frac{1}{\Lambda_\eta}}.
\end{align}
At this classical dimensional crossover, the logarithmic increase of
(\ref{CorrelationLengthExplicit2dQC-AFMFM}) with increasing correlation length $\xi$ is
cutoff. This allows, in particular, a solution for the phase boundary, see below. However, at a higher temperature
\begin{align} \label{CrossT-FMAF}
T_x \sim \frac{\Lambda_\eta^2}{u},
\end{align}
$\Tcl(r)$ enters the Ginzburg regime and the classical dimensional crossover and, as a consequence, the expression for the phase boundary becomes non-perturbative.

{\it Non-critical Fermi liquid regime, $T \ll \Lambda_\eta^3$ and $r_{2d} \ll \Lambda_\eta^2$:}
The 2d critical behavior crosses over into a 2d non-critical Fermi
liquid behavior below a temperature $T \sim \Lambda_\eta^3$ or, as a function of the
control parameter at $r \sim \Lambda^2_\eta$, see the shaded lines in
Fig.~\ref{fig:pd-FMAF}. In this intermediate regime, the correlation length has the form,
\begin{align} \label{CorrelationLengthNonCritical-AFMFM}
\xi^{-2} =
r_{2d} + \mathcal{C}_1 u \frac{T^2}{\Lambda_\eta^3},
\end{align}
where the numerical factor $\mathcal{C}_1$ 
is non-universal, i.e., depends on our modeling of the dimensional crossover. It turns
out that this 2d Fermi-liquid background still dominates thermodynamics down to a
temperature and control parameter scale $\Lambda_\eta^4$ where finally the 3d critical
behavior takes over.

\subsubsection{Correlation length in the 3d regime, $T \ll \Lambda_\eta^4$ and $r_{3d} \ll \Lambda_\eta^4$}

{\it Fermi-liquid regime, $T\ll r_{3d}$:}
The contribution from the 3d fluctuations shift the value for the control parameter
\begin{align}
\xi^{-2} = r_{3d} \equiv r_{2d} + \mathcal{C}_2 u \Lambda^2_{\eta} + \frac{\pi (N+2)}{144} u \frac{T^2}{\Lambda_\eta r_{3d}^{1/2}}
\end{align}
where $\mathcal{C}_2$ is a constant dependent on the employed cutoff structure. The
temperature dependence is again of Fermi liquid type. Comparing the temperature corrections to the correlation length in the various Fermi liquid regimes, one obtains the two control parameter crossover scales, $\Lambda_\eta^4$ and $\Lambda_\eta^2$, that are shown in Fig.~\ref{fig:pd-FMAF}.

{\it Quantum critical regime, $T\gg r_{3d}$:}
In the quantum critical regime we get
\begin{align} \label{CorrelationLengthExplicit3dQC-AFMFM}
\xi^{-2} =
r_{3d} + (N+2) \frac{\zeta(3/2)}{24 \sqrt{2 \pi}} u \frac{T^{3/2}}{\Lambda_\eta}.
\end{align}
A comparison of this temperature dependence with the one of
Eq.~(\ref{CorrelationLengthNonCritical-AFMFM}) yields the quantum dimensional crossover
temperature, $T \sim \Lambda_\eta^4$.

\subsubsection{Phase boundary}

In the limit of vanishing correlation length,
Eq.~(\ref{CorrelationLengthExplicit2dQC-AFMFM}) yields for $T_c <T_x$, see Eq.~(\ref{CrossT-FMAF}), the following expression for the
phase boundary within the quantum critical 2d regime
\begin{align}
r_{2d} = - \frac{N+2}{24 \pi} u T_c \log \frac{T_c^{2/3}}{ \Lambda^2_\eta}.
\end{align}
It varies (up to logarithmic corrections) linearly with $T$. In the intermediate
non-critical Fermi liquid regime, the phase boundary behaves as
$r_{2d} \sim - u T_c^2/\Lambda_\eta^3$, before crossing over into the 3d regime.
There, we use expression (\ref{CorrelationLengthExplicit3dQC-AFMFM}) to obtain
\begin{align}
r_{3d} = -  (N+2) \frac{\zeta(3/2)}{24 \sqrt{2 \pi}} u \frac{T_c^{3/2}}{\Lambda_\eta}.
\end{align}
Similar to Sec.~\ref{sec:PhaseBoundaryAFM-AFM}, the extrapolation of the phase boundary
within the 2d regime towards $T=0$ leads to an error in the estimate for position of the
quantum critical point of
\begin{align}\label{ErrorExtrapolation2}
\Delta r = \left.r_{3d}\right|_{r_{2d}=0} \sim u \Lambda^2_{\eta}.
\end{align}

\subsection{Thermodynamics}

\begin{figure}
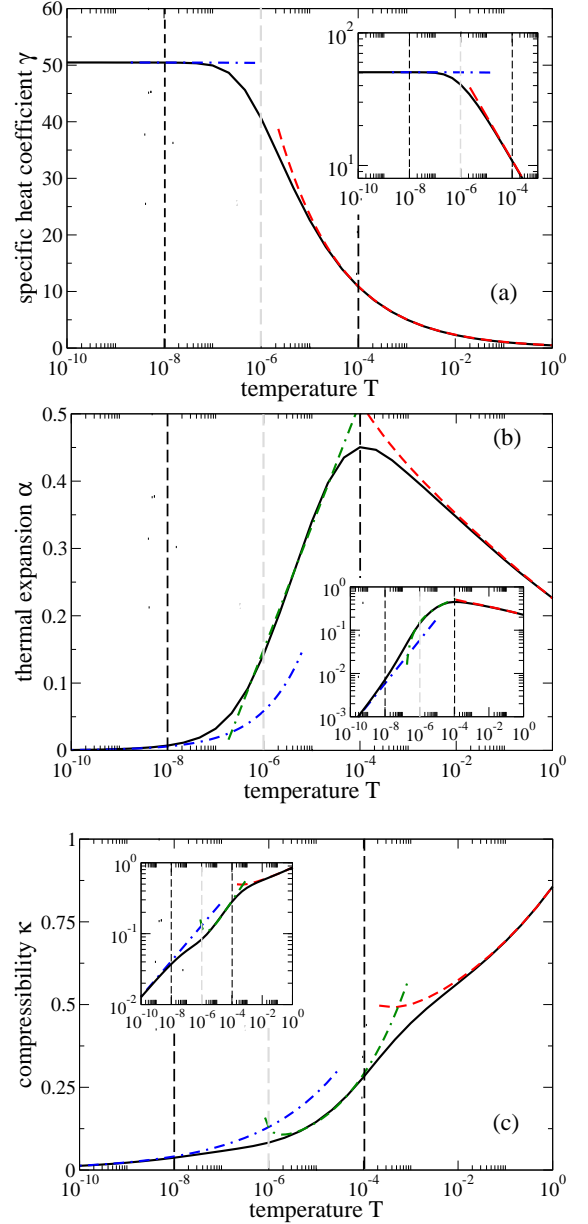

\includegraphics[width=0.41\textwidth]{fig6a.eps}
\vspace{1em}
\includegraphics[width=0.41\textwidth]{fig6b.eps}
\vspace{1em}
\includegraphics[width=0.41\textwidth]{fig6c.eps}
\caption{(color online)
Crossover behavior as in Fig.~\protect\ref{fig:2af3af}, but
now for the 2d FM -- 3d AFM case.
The vertical dashed lines denote the three crossover scales $\Tcl > \Lambda^3_\eta >
\Lambda^4_\eta$, see Fig.~\protect\ref{fig:pd-FMAF}.
$\Tcl$ denotes the classical dimensional crossover where $\xi \sim 1/\Lambda_\eta$;
$T \sim \Lambda^3_\eta$ separates the 2d critical regime from the 2d non-critical Fermi-liquid
regime; and below $T \sim \Lambda_\eta^4$ 3d quantum critical behavior sets in.
The chosen parameters are $\Lambda_\omega = \Lambda = 20$,
$\Lambda_\eta = 0.01$, $u=1$ and $N=3$.
Panel (a): Specific-heat coefficient $\gamma$,
with the asymptotics given by
Eq.~(\ref{SpecificHeatQCR-FMAF-2d}) (red/dashed) for the high-temperature 2d regime and by
Eq.~(\ref{SpecificHeatQCR-d>z}) 
(blue/dash-dot) for the low-temperature 3d regime.
Panel (b): Thermal expansion $\alpha$, showing a maximum at the classical dimensional crossover.
The asymptotics at high (red/dashed) and intermediate $T$ (green/dash-dash-dot) are given
by (\ref{ThermExpQCR-FMAF-2d-explicit}), and at low $T$ (blue/dash-dot) by Eq.~(\ref{ThermExpQCR-d>2})
The inset shows that the asymptotic $\sqrt{T}$ behavior only sets in for $T < \Lambda_\eta^4 = 10^{-8}$.
Panel (c): Compressibility $\kappa$, with the asymptotics in
Eq.~(\ref{CompressQCR-FMAF-2d}) (red/dashed and green/dash-dash-dot) and Eq.~(\ref{Compressibility3dQCAFM-AFM}) (blue/dash-dot)
showing the $T^{1/4}$ behavior that again only sets in for $T < \Lambda_\eta^4$.
}
\label{fig:2f3af}
\end{figure}

Thermodynamic quantities follow from the expression (\ref{CriticalFreeEnergy}) for the free
energy. A numerical solution for the specific heat, thermal expansion and compressibility
in the quantum critical regime is shown in Fig.~\ref{fig:2f3af}. In the following, an
analysis of the asymptotic behavior is presented.

\subsubsection{Thermodynamics in the 2d regime, $T \gg \Lambda_\eta^4$ or $r_{2d} \gg \Lambda_\eta^4$
}

{\it Fermi-liquid regime, $T\ll r_{2d}^{3/2}$ and $r_{2d} \gg \Lambda_\eta^2$:}
In the 2d Fermi-liquid regime, evaluating the leading behavior is straightforward. For
the specific-heat coefficient and thermal expansion we obtain
\begin{align}
\gamma_{\rm cr} = \frac{N \pi}{12} \frac{1}{\sqrt{r_{2d}}},
\qquad
\alpha_{\rm cr} = \frac{N \pi}{24} \frac{T}{r_{2d}^{3/2}}.
\end{align}
This gives a universal Gr\"uneisen parameter\cite{Zhu03}
\begin{align}
\Gamma_{\rm cr} = \frac{1}{2} \frac{1}{r_{2d}}.
\end{align}
For the critical part of the compressibility we find
\begin{align}
\kappa_{\rm cr} = - \frac{N}{8\pi} \sqrt{r_{2d}}.
\end{align}

{\it Quantum critical regime, $T\gg r_{2d}^{3/2}$ and $T \gg \Lambda_\eta^3$:}
In the 2d quantum critical regime, the specific heat is given by
\begin{align} \label{SpecificHeatQCR-FMAF-2d}
\gamma_{\rm cr} = \frac{N}{6\pi} \Gamma\left(\frac{8}{3}\right) \zeta\left(\frac{5}{3}\right) T^{-1/3}.
\end{align}
The thermal expansion depends logarithmically on temperature
\begin{align} \label{ThermExpQCR-FMAF-2d}
\alpha_{\rm cr} =
\frac{N}{8\pi} \log\frac{T^{2/3}}{{\rm max}\{\xi^{-2},\Lambda^2_\eta\}} + {\rm const}\,,
\end{align}
with the correlation length given in Eq.~(\ref{CorrelationLengthExplicit2dQC-AFMFM}).
Using its explicit form, it becomes clear that the thermal expansion has a maximum at the
dimensional crossover, $\xi^{-1} \sim \Lambda_\eta$, that occurs at a temperature $\Tcl$
and is associated with the classical Matsubara zero mode:
\begin{align} \label{ThermExpQCR-FMAF-2d-explicit}
\alpha_{\rm cr} \sim \left\{
\begin{array}{ll}
 \log\frac{1}{u T^{1/3}}
& {\rm if}\quad T \gg \Tcl 
\\[0.5em]
  \log\frac{T^{2/3}}{\Lambda^2_\eta} + {\rm const.}
& {\rm if}\quad T \ll \Tcl 
\end{array}
\right. .
\end{align}
In the quantum critical 2d regime, the Gr\"uneisen parameter shows the asymptotic behavior
\begin{align}
\Gamma_{\rm cr} \sim \frac{1}{T^{2/3}},
\end{align}
where we omitted logarithmic corrections that depend on the effective dimensionality of
the classical fluctuations, see Eq.~\eqref{ThermExpQCR-FMAF-2d-explicit}.

Similarly, the compressibility also exhibits an additional dimensional crossover at
$\Tcl$ as it is determined by the classical fluctuations
\begin{align} \label{CompressQCR-FMAF-2d}
\kappa_{\rm cr} =
\left\{
\begin{array}{ll}
\frac{N}{8\pi} \frac{T}{\xi^{-2}}
& {\rm if}\quad T \gg \Tcl 
\\[0.5em]
\frac{N}{16} \frac{T}{\Lambda_\eta \xi^{-1}}
& {\rm if}\quad T \ll \Tcl 
\end{array}
\right.
\end{align}
Using Eq.~(\ref{CorrelationLengthExplicit2dQC-AFMFM}) for the correlation length we
obtain the asymptotic behavior for the compressibility at criticality, $r \approx r_{2d}
= 0$,
\begin{align}
\label{CompressQCR-FMAF-2d-explicit}
\kappa_{\rm cr} \sim
\left\{
\begin{array}{ll}
\frac{1}{u \log{\frac{1}{T}}}.
& {\rm if}\quad T \gg \Tcl 
\\[0.5em]
\frac{\sqrt{T}}{\Lambda_\eta \sqrt{u}}
& {\rm if}\quad T \ll \Tcl 
\end{array}
\right. .
\end{align}
Note that $\kappa$ depends singularly on the dangerously irrelevant interaction $u$.

The results \eqref{ThermExpQCR-FMAF-2d-explicit} and \eqref{CompressQCR-FMAF-2d-explicit}
show that not only the 2d quantum critical regime above $\Tcl$, but
also the intermediate 2d quantum 3d classical critical regime below $\Tcl$
displays a well-defined universal asymptotic behavior.

{\it Non-critical Fermi liquid regime, $T \ll \Lambda_\eta^3$ and $r_{2d} \ll \Lambda_\eta^2$:}
In the intermediate non-critical regime, thermodynamic quantities display a Fermi-liquid
form:
\begin{align}
\gamma &= \frac{1}{2\Lambda_\eta} - \mathcal{C}_3 \frac{T^2}{\Lambda_\eta^7},\qquad
\alpha = \mathcal{C}_4 \frac{T}{\Lambda_\eta^{3}}
\end{align}
with constants $\mathcal{C}_3$ and $\mathcal{C}_4$ being non-universal.
The compressibility is determined by 3d classical fluctuations,
\begin{align}
\kappa_{\rm cr} = \frac{N}{16} \frac{T}{\Lambda_\eta \xi^{-1}},
\end{align}
where the correlation length $\xi$ has the form (\ref{CorrelationLengthNonCritical-AFMFM}).

\subsubsection{Thermodynamics in the 3d regime, $T \ll \Lambda_\eta^4$ and $r_{3d} \ll \Lambda_\eta^4$
}

The thermodynamics in the 3d regime resembles the one discussed in
Sec.~\ref{sec:ThermodynamicsAFM-AFM}. The background contribution for the specific heat
coefficient in Eq.~(\ref{SpecificHeatFL2d}) is, however, now determined by the 2d FM
fluctuations, $\gamma - \gamma_{\rm cr} = \frac{1}{2\Lambda_\eta}$.

\subsection{Uniform susceptibility}
\label{Sec:susceptibility}

\begin{figure}
\includegraphics[width=0.48\textwidth]{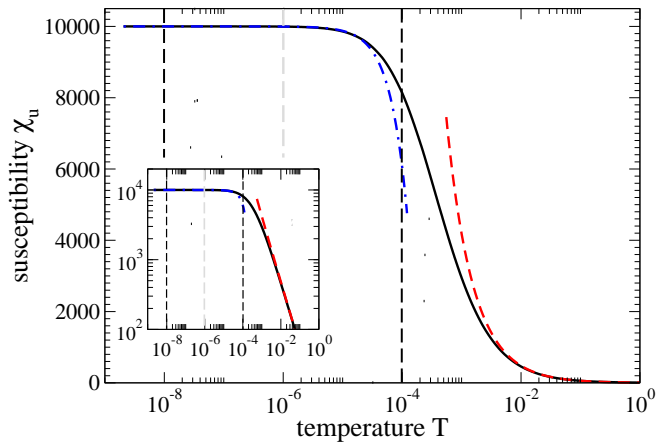}
\caption{
Behavior of the uniform susceptibility $\chi_{\rm u}$ for $r=0$, i.e., in the quantum critical regime of
the 2d FM -- 3d AFM crossover scenario. The chosen parameters are $\xi_Q^{-2} = 0.0001$,
$\Lambda_\omega = \Lambda = 20$, $\Lambda_\eta = 0.01$, $u=1$ and $N=3$. The vertical
dashed lines represent the crossover scales $T_Q = \xi_Q^{-2} \sim \Lambda_\eta^2 < \Lambda_\eta^3 < \Lambda_\eta^4$. The dashed/red line is the
2d high-$T$ asymptote (\ref{Sus2d}), and the dashed-dot/blue line is the 3d low-$T$
asymptote (\ref{Sus3d}).
}
\label{fig:2f3afchi}
\end{figure}

A particular property of the 2d FM -- 3d AFM crossover is that the susceptibility,
measuring the response to a uniformly applied magnetic field, is critical within the 2d
regime, whereas it becomes non-critical at the dimensional crossover.
We consider the static limit and study its temperature dependence
\begin{eqnarray} \label{SusceptibilityFMAF}
\chi_{\rm u}(T) \equiv \chi (T,\Omega_n=0,k_\parallel =0, k_z = Q_z)= \frac{1}{\xi^{-2} + \xi^{-2}_Q } \; . \nonumber \\
\end{eqnarray}
As before, the wavevector $\bf k$ is measured relative to the (3d) ordering wavevector
$\bf Q$.
$\xi^{-2}_Q$ is a constant proportional to the hopping $t'$ between the 2d planes,
$\xi^{-2}_Q \propto t' [1-\cos(Q_z a)]$.
From the discussion in Sec.~\ref{sec:action} it follows that
$\xi^{-2}_Q \sim \Lambda^2_\eta$.
The temperature behavior is fully accounted for by the $T$ dependence
of the correlation length $\xi$. In the quantum critical 2d regime, i.e., at high
temperature, we can neglect the constant contribution $\xi^{-2}_Q$, and we obtain with the
help of (\ref{CorrelationLengthExplicit2dQC-AFMFM}) for the asymptotic behavior
\begin{eqnarray} \label{Sus2d}
\chi_{\rm u}(T) \sim \frac{1}{T \log \frac{1}{T}}.
\end{eqnarray}
This divergence is cut off below a scale $T_Q = \xi_Q^{-2} \sim  \Lambda^2_\eta$,
i.e. essentially at the  classical dimensional crossover line.
For $T \ll T_Q$, the constant contribution from the $t'$ dispersion dominates,
$\chi_{\rm u} =$ const.
In the 3d quantum critical regime, the temperature dependence of the correlation
length (\ref{CorrelationLengthExplicit3dQC-AFMFM}) leads to the $T$-dependent correction
$\chi_{\rm u} = $const $+\delta \chi_{\rm u}(T)$,
\begin{eqnarray}\label{Sus3d}
\delta\chi_{\rm u} \sim - T^{3/2} \; ,
\end{eqnarray}
in agreement with the results of Ref.~\onlinecite{ioffe95}.
The result of a numerically evaluation of the uniform susceptibility
(\ref{SusceptibilityFMAF}) is shown in Fig.~\ref{fig:2f3afchi}.
The lower two crossover temperatures leave essentially no trace in $\chi_{\rm u}$,
as the regimes only differ in the small thermal correction to the large
$\chi_{\rm u}(T\!=\!0)$.


\section{Application to heavy-fermion metals}
\label{sec:exp}

Our results are of potential relevance to layered nearly magnetic metals
where indications for two-dimensional criticality have been found.
In this section, we discuss the cases of the
heavy-fermion materials \CeAu, \YbRhSi, and \CeCoIn.
While \CeCoIn\ possesses a layered lattice structure and should naturally display a
dimensional crossover, for \CeAu\ and \YbRhSi\ there is only empirical evidence
(discussed below) for quasi-2d critical behavior, with no obvious
microscopic reason.
We note that, in all three cases, ingredients beyond the
LGW spin fluctuation theory may be important for a full
understanding of the critical behavior.

We refrain from discussing strongly correlated transition-metal compounds
with layered structure, such as high-temperature superconducting cuprates.
In these materials, a standard LGW approach alone cannot be expected to capture
the relevant physics due to the proximity to the half-filled Mott insulator.

\subsection{\CeAu}

CeCu$_6$ is a paramagnetic heavy Fermi liquid, which can be driven
into an antiferromagnetic metallic phase by Au substitution.\cite{hvl,hvl94}
In \CeAu, the quantum critical point is located at $x_c \approx 0.1$.
For $x>x_c$, the AF order is known to be 3d, which implies non-vanishing magnetic
couplings in all spatial directions.
In contrast, neutron scattering in quantum-critical \CeAu\ has provided direct
evidence for a quasi-2d antiferromagnetic fluctuation spectrum.\cite{Stockert98}
This suggests that a dimensional crossover occurs near the quantum critical
point, which, however, has not been experimentally identified to date.
Let us therefore apply our results from Sec.~\ref{sec:2af3af}
in order to search for experimentally measurable consequences of
a dimensional crossover from 2d AFM to 3d AFM.

First, there is the location of the QCP itself.
Experimentally, the phase boundary appears to be linear, $\TN \propto (x-0.1)$,
in other words, the data points marking the finite-temperature phase
transition can be linearly extrapolated to a putative QCP at $x_c=0.1$.
From the consideration in Sec.~\ref{sec:2af3af} it is clear that
the asymptotic behavior of the phase boundary should be $\TN \propto (x-x_c)^{2/3}$,
which suggests that the true $x_c > 0.1$.
Hence, samples with $x=0.1$ may not be located at the quantum critical composition.
As a consequence, the system would be located outside the 3d pocket of the
phase diagram in Fig.~\ref{fig:asd}, this could then explain why no signatures of
3d spin fluctuation were found in the neutron scattering experiment of
Ref.~\onlinecite{Stockert98}.
Unfortunately, concrete predictions are problematic due to the presence of logarithmic
corrections in the $d\!=\!z\!=\!2$ theory for the 2d AFM:
In fact, in the 2d regime the phase boundary should follow
$\TN\log\TN \propto (x-x_c)$, Eq.~\eqref{PhaseBoundary2d} --
this makes a linear extrapolation ambiguous.
Experimentally, such logarithmic corrections are difficult to extract.

Second, the dimensional crossover should be manifest in thermodynamics. Whereas the
crossover signatures in the specific heat are weak, the thermal expansion and the
compressibility show a pronounced step-like behavior in the quantum critical regime of
the $2d-3d$ AFM crossover, see Fig.~\ref{fig:2af3af}. The latter two are therefore more
appropriate to detect a dimensional crossover in the critical spin-fluctuation spectrum.

Thus, we propose to search for the dimensional crossover in \CeAu\ by
(i) detecting the change in behavior of the thermal expansion at criticality and (ii) by looking for deviation of  the phase boundary from linear behavior by employing pressure tuning of, e.g., an $x=0.2$ sample. From the available data, the dimensional crossover temperature is likely below 100 mK.
We also note that the presence of quenched chemical disorder may
modify the behavior near criticality at very low temperatures.\cite{tv}

\subsection{\YbRhSi}

The heavy-fermion material \YbRhSi\ displays a phase transition
at 70 mK.\cite{trovarelli}
The low-temperature ordered phase is believed to be antiferromagnetic,
although confirming neutron scattering data are not available
to date.
The ordering temperature can be suppressed by applying a small field, resulting
in a field-driven QCP at $B=0.06$ T in the $ab$ plane and
0.66 T along the $c$ axis.
The ordering temperature can also be suppressed by Ge doping:
\YbRhSG\ with $x=5$\% seems to order at only 20 mK in zero field.\cite{custers03}
The critical properties of \YbRhSi\ are inconsistent with the predictions of
LGW theory for 3d AFM spin fluctuations.
As for \CeAu, it has been speculated that the Kondo effect breaks down
at quantum criticality -- this idea received support from Hall effect
measurements which indicate a pronounced change in the low-temperature
Hall coefficient across the critical field.\cite{paschen04}

Remarkably, \YbRhSi\ appears to be almost {\em ferro}magnetic.
This is particularly striking for \YbRhSG\ where
the uniform susceptibility follows $\chi_{\rm u}(T) \propto T^{-0.6}$ above 0.3 K.\cite{gegenwart05}
In addition, the unexpected observation of an ESR signal below
the Kondo temperature\cite{sichel} has been related to strong
ferromagnetic correlations.\cite{krellner}

It is therefore worth discussing which properties of
\YbRhSi\ appear consistent with the scenario of a crossover from
2d FM to 3d AFM, as would arise in a system of ferromagnetic layers with weak
antiferromagnetic inter-layer coupling.

Interestingly, thermodynamic measurements are partially consistent with
2d FM criticality, but {\em below} 0.3 K.
The specific heat follows $C(T)/T \propto T^{-0.3}$ below 0.3\,K,\cite{custers03}
and the Gr\"uneisen ratio diverges as $\Gamma(T) \propto T^{-0.7}$ below 0.6\,K
(Ref.~\onlinecite{kuechler03}) -- these
two exponents are close to the values $-1/3$ and $-2/3$ expected
for 2d FM fluctuations.
In addition, the temperature--field scaling observed in \YbRhSG\ with $x=5\%$
(Ref.~\onlinecite{gegenwart05b}) is in accord with ferromagnetic criticality,
provided that one interprets $(B-B_c)$ (where $B_c=0.027$ T is tiny)
as the field conjugate to the order parameter.
However, other observations in \YbRhSi\ appear inconsistent with this
idea of 2d FM criticality:\cite{gegenwart07} for example, the fractional exponent observed in the $T$ dependence of the uniform susceptibility $\chi_{\rm u}(T)$ cannot be easily explained with this scenario.

In summary, the physics of \YbRhSi\ cannot be explained in a straightforward
manner in terms of near-critical 2d FM fluctuations (which turn to 3d
antiferromagnetism at lowest temperatures) alone.
However, the experiments, showing critical ferromagnetic fluctuations
which are cutoff only at very low temperatures, hint that a
dimensional crossover of the type considered here may be important.
Further investigations of \YbRhSi\ samples with Ir or Co doping\cite{gegpriv}
will shed more light on the role of the various crossover scales in this
interesting material.

\subsection{CeCoIn$_5$}

The compounds CeMIn$_5$ (M = Co,Rh,Ir, also dubbed 115-compounds),
unify a variety of fascinating heavy-fermion phenomena:
CeCoIn$_5$ and CeIrIn$_5$ are (likely unconventional) superconductors with
$\Tc = 2.3$ K and 0.4 K, respectively,
whereas CeRhIn$_5$ is an antiferromagnetic metal with $\TN \approx 3.6$ K.
Transitions between these ordered phases may be tuned
using pressure, chemical substitution, or magnetic field.
In contrast to most other heavy-fermion materials, CeMIn$_5$ is
quasi two-dimensional, i.e., its lattice structure consists
of weakly coupled layers.
Consequently, a dimensional crossover scenario should naturally apply.

A particularly interesting transition occurs in \CeCoIn\ upon
application of a magnetic field.\cite{bianchi03b}
Superconductivity survives up to a critical field of $H_{c2} \approx 4.95$ T.
Normal-state properties near $H_{c2}$ are suggestive of
quantum critical behavior:
both specific heat and resistivity display non-Fermi liquid temperature
dependencies at $H_{c2}$,
and the $A$ coefficient of the resistivity diverges upon approaching $H_{c2}$
from above.
These features have been interpreted as signatures of an antiferromagnetic
quantum critical point at (or close to) $H_{c2}$, with the ordered phase
for $H<H_{c2}$ being suppressed by the onset of superconductivity.
By applying hydrostatic pressure, the two phenomena -- superconducting $H_{c2}$
transition and apparent magnetic quantum criticality -- can be
separated, i.e. $H_{c2}$ decreases faster than the quantum critical field,
suggesting that the two phenomena are not related.\cite{ronning06}

Recently, thermodynamic properties of CeCoIn$_5$ near $H_{c2}$ have been
studied in more detail.\cite{donath}
At $H=5$ T, signatures of a crossover at $T^\ast \approx 0.3$ K between two different
singular behaviors were identified.
This is best visible in the thermal expansion,
which was found to follow $\alpha(T)/T \propto 1/T$ in the temperature range
0.5 K $< T <$ 6 K, whereas $\alpha(T)/T$ is consistent with $1/\sqrt{T}$
for 0.1 K $< T <$ 0.3 K.
The thermodynamic data below $T^\ast \approx 0.3 K$ have been argued to be consistent
with the predictions of the LGW theory for 3d AF spin fluctuations.
(Note that the scale $T^\ast$, below which 3d LGW behavior is seen,
is shifted up to 1.4 K in doped CeCoIn$_{5-x}$Sn$_x$.)
For $T>T^\ast$, the thermal expansion in CeCoIn$_5$ seems consistent
with 2d AF spin fluctuations of LGW type.
However, the authors of Ref.~\onlinecite{donath} have argued that the behavior
of the Gr\"uneisen parameter, being is somewhat reminiscent to that of \YbRhSi,
instead suggests non-LGW criticality (which nevertheless may
arise from 2d critical magnetism).

Taken together, the data indicate a crossover in critical behavior at
$T^\ast\approx 0.3$ K -- this is also supported by resistivity
measurements which show $\rho(T)-\rho(0) \propto T^{3/2}$ below $T=0.2$ K
at $H=5$ T,\cite{paglione06} consistent with 3d LGW behavior.
Although the nature of the critical behavior above $T^\ast$ is not
fully understood -- the presence of multiple crossover scales\cite{paglione06}
complicates the analysis of the data -- the interpretation of $T^\ast$ as a
dimensional crossover scale is suggestive.
The role of Sn doping in shifting this crossover scale is unclear at present;
in a situation with geometric frustration of inter-plane magnetism one might
speculate that disorder partially relieves frustration.
It would be worthwhile to investigate the magnetic excitations, e.g.,
of AF ordered variants of CeMIn$_5$ by neutron scattering,
in order to determine the magnetic bandwidths in the directions parallel and
perpendicular to the CeIn planes.


\section{Conclusion}
\label{sec:concl}

We have studied the dimensional crossover of magnetic fluctuations
in nearly quantum critical metals.
Motivated by experiments on heavy-fermion systems, we have concentrated
on the crossover from 2d FM or AFM fluctuations at elevated energies
to 3d AFM fluctuations at low energies.
Applying the standard Landau-Ginzburg-Wilson approach, we
have obtained relevant crossover energy scales as well as
crossover functions describing thermodynamic observables.

The anisotropy in the spin-fluctuation spectrum leads to a dimensional crossover from 2d to 3d upon approaching criticality. We have found two types of dimensional crossover scales. Upon reducing temperature at criticality, there is a first dimensional crossover associated with the classical fluctuations and, at a lower temperature, a second crossover where the quantum fluctuations change their character from 2d to 3d, see Figs.~\ref{fig:pd-AFAF} and \ref{fig:pd-FMAF}. In particular, there is an extended intermediate temperature regime where 2d quantum fluctuations coexist with 3d classical fluctuations resulting in new power laws. For the 2d FM - 3d AFM crossover, there exists even a further sub-regime, where non-critical Fermi liquid behavior intervenes between the critical 2d and 3d regimes. We have found that the thermal expansion and the compressibility are well suited to detect a dimensional crossover of critical magnetic fluctuations. However, the existence of several dimensional crossover scales makes the experimental identification of asymptotic power laws particularly difficult.


\acknowledgments

We thank S. Florens, P. Gegenwart, and P. W\"olfle
for discussion and collaboration on related work.
This research was supported by the DFG through the SFB 608,
the Research Unit FG 960 ``Quantum Phase Transitions'',
and grant FR 2627/1-1,
and by the NSF through grant
DMR-0757145.


\appendix

\section{Lindhard function for a Fermi surface with cylindrical symmetry}
\label{appen:lindhard}

\begin{figure}
\includegraphics[width=0.47\textwidth]{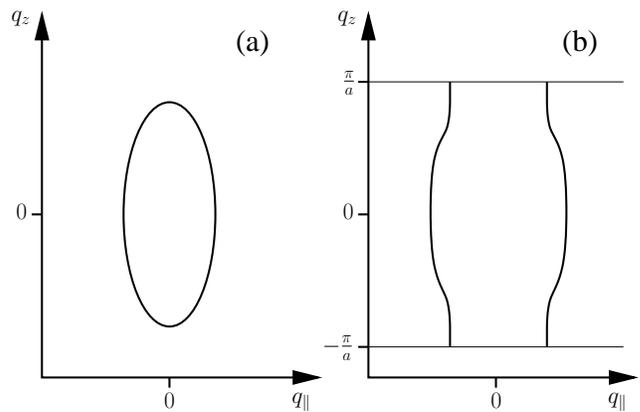}
\caption{
Sketch of a cut through a closed (a) and open (b) 3d anisotropic Fermi surface cylindrically symmetric with respect to the $q_z$-axis; $q_z$ is the out-of-plane, and $q_\parallel$ is the in-plane momentum.}
\label{FermiSurface}
\end{figure}

We evaluate the Landau damping of ferromagnetic fluctuations deriving from the Lindhard
function for an anisotropic Fermi surface with cylindrical symmetry. The Lindhard function at $T=0$ is given by
\begin{align}
\Pi(k,i\omega_n) = - \int \frac{d\epsilon}{2\pi} \sum_q \frac{1}{i \epsilon -\varepsilon_q}
\frac{1}{i \epsilon + i \omega_n -\varepsilon_{q+k}}.
\end{align}
%
%
In the following, we distinguish between a closed and an open anisotropic Fermi surface.

\subsection{Closed Fermi surface}

Here, we consider a closed anisotropic Fermi surface as, e.g., depicted in Fig.~\ref{FermiSurface}a.
Introducing the density of states and integrating over $\epsilon$, the dynamic part of the polarization can be expressed as an integral over the (closed) 3d Fermi surface
\begin{align}
\Pi_{\rm dyn}(k,i\omega_n) = - i \omega_n  \int \frac{d \hat{q}}{4\pi}
\frac{\nu_{\hat{q}}}{i \omega_n  - {\bf v}_{\hat{q}} {\bf k}},
\end{align}
For an anisotropic Fermi surface, the density of states $\nu_{\hat{q}}$ and the Fermi velocity ${\bf v}_{\hat{q}}$ depend on the orientation of the fermionic wavevector, $\hat{q}$. We model the anisotropy of the Fermi surface with an in-plane Fermi velocity $v_F$ that
differs from its z-component $\eta_F v_F$. A large anisotropy with a quasi
two-dimensional Fermi surface is obtained in the limit of small $\eta_F$. So we get
\begin{align}
\Pi_{\rm dyn}(k,i\omega_n) &=
 - \frac{1}{4\pi}\, i \omega_n  \int^\pi_0 d\theta \int_0^{2\pi} d\phi
\\\nn&\times
\frac{\sin\theta\, \nu(\cos^2\theta)}{i \omega_n  - v_F (k_\parallel \sin \theta \cos \phi + \eta_F k_z \cos
\theta)}.
\end{align}
We further assumed that the density of states only depends on $\cos^2\theta$ with the
azimuthal angle $\theta$ of the Fermi momentum. In the limit of small $\omega_n$, this
simplifies to
\begin{align} \label{PolResult}
\Pi_{\rm dyn}(k,i\omega_n) &=
- \frac{\pi}{2} \langle \nu \rangle \frac{|\omega_n|}{v_F \sqrt{k_\parallel^2 + (\eta_F k_z)^2}}
\end{align}
where $\langle \nu \rangle$ is an angular average of the density of states that smoothly depends on momenta
\begin{align} \label{AvDOS}
\langle \nu \rangle =
\frac{2}{\pi} \int_0^1 \frac{dx}{\sqrt{1-x^2}}\,  \nu\left(\frac{k^2_\parallel x^2}{k_\parallel^2+\eta_F^2
k_z^2}\right).
\end{align}
Generally, the damping of bosonic modes with momentum ${\bf k}$ is caused by
particle-hole excitations close to the part of the Fermi surface that is tangential to
${\bf k}$. This is reflected in the momentum dependence of $\langle \nu \rangle$. For the
modes with vanishing in-plane momentum, $k_\parallel = 0$, only the part of the Fermi
surface with azimuthal angle $\theta = \pi/2$ is involved in the damping processes, such
that $\langle \nu \rangle = \nu(0)$. On the other hand, the damping of modes with
vanishing $k_z = 0$ can occur by exciting particle-hole pairs at any part of the Fermi
surface and in this case $\langle \nu \rangle$ is a true average over the angle-dependent
density of states.

Thus we obtain for the damping function $\Gamma_k$, see
Eq.~(\ref{AnistropicSusceptibility}), for an anisotropic Fermi surface
\begin{align} \label{lind_res}
\Gamma_k = \frac{2 v_F}{\pi \langle\nu\rangle}\sqrt{k_\parallel^2 + (\eta_F k_z)^2}.
\end{align}
In the 2d regime, damping is dominated by the in-plane momentum, $\Gamma_k \sim
|k_\parallel|$, and we obtain the expression advertised in Eq.~(\ref{damping2}).

\subsection{Open Fermi surface}

In the limit of a quasi two-dimensional Fermi liquid, the Fermi surface opens as depicted in Fig.~\ref{FermiSurface}b. If we neglect the warping of the Fermi-surface cylinder along the momentum $q_z$ direction, we end up with a Lindhard function of an effectively 2d Fermi system that is, in particular, independent of the longitudinal momentum $k_z$. Its dynamic part then has the form
\begin{align} \label{Pol2dLimit}
\Pi_{\rm dyn}(k_\parallel ,i\omega_n) = - \nu_\parallel \frac{|\omega_n|}{v_F k_\parallel} \sum_{q_z}
\end{align}
with the 2d density of states $\nu_\parallel$. The integration over the longitudinal momentum $q_z$ just yields a multiplicative factor.

If we take the warping into account, we obtain instead  the same limiting behavior as that of expression (\ref{PolResult}). In the limit $\eta_F k_z \ll k_\parallel$, where $\eta_F$ is again a small parameter representing the strong anisoptropy of the open Fermi surface, the result is essentially unchanged from Eq.~(\ref{Pol2dLimit}) except that $\nu_\parallel$ is replaced by an averaged density of states. In the other limit of small in-plane momentum $\eta_F k_z \gg k_\parallel$, the warping in Fig.~\ref{FermiSurface}b leads to a $k_z$ dependence of the dynamic part of the polarization resulting from  particle-hole excitations now concentrated at the center, $q_z \approx 0$, and at the edges, $q_z \approx \pm \pi/a$, of the Brillouin zone, where the Fermi surface is parallel to $k_z$.



\begin{thebibliography}{99}
\bibitem{intnote}
We assume that the
singular terms in the LGW expansion discussed in Refs.~\onlinecite{bkv,chub2dafm}
are negligible in the 2d intermediate-energy regime of the layered system under
consideration.
(The singularities are not present in the asymptotic 3d AFM regime.)
As the singularities are interaction-generated, microscopic details become
relevant here. E.g. close to nesting, AF criticality will result already from small
interactions, such that the non-LGW regime is expected to be pushed to very small
energy scales.
\end{thebibliography}
\end{document}